\documentclass{aa}  

\usepackage{xcolor}
\usepackage{graphicx}
\usepackage{multirow}
\usepackage{subcaption}
\usepackage{caption}
\usepackage{comment}
\usepackage{threeparttable}
\usepackage{array}
\usepackage{placeins}

\usepackage{ulem}
\RequirePackage{color}

\usepackage[switch]{lineno}

\usepackage{hyperref}
\usepackage{graphicx}
\usepackage{txfonts}

\newcommand{\Lya}{\mbox{Ly-$\alpha$}}

\begin{document}

   \title{A view of the evolution of a CME and the associated wave-trains at high spatial and temporal resolution.}

   \titlerunning{short title}

   \author{G. Russano\inst{\ref{INAF-OAC}}
        \and
        Y. De Leo\inst{\ref{INAF-OACt},\ref{UniGRAZ}}
        \and
        F. Frassati\inst{\ref{INAF-OATo}}
        \and
        G. Jerse\inst{\ref{INAF-OATs}}
        \and
        V. Andretta\inst{\ref{INAF-OAC}}
        \and
        H. Cremades\inst{\ref{CONICET}}
        \and
        M. Temmer\inst{\ref{UniGRAZ}}
        \and
        S. Mancuso\inst{\ref{INAF-OATo}}
        \and
        L. Abbo\inst{\ref{INAF-OATo}}
        \and
        A. Burtovoi\inst{\ref{INAF-OATo},\ref{UniFi}}
        \and
        F. Landini\inst{\ref{INAF-OATo}}
        \and
        M. Pancrazzi\inst{\ref{INAF-OATo}}
        \and
        M. Romoli\inst{\ref{UniFi},\ref{INAF_Arcetri}}
        \and
        C. Sasso\inst{\ref{INAF-OAC}}
        \and
        R. Susino\inst{\ref{INAF-OATo}}
        \and 
        M. Uslenghi\inst{\ref{INAF-IASF}}
          }

   \institute{
        National Institute for Astrophysics (INAF), Astronomical Observatory of Capodimonte, Napoli, Italy\\
              \email{giuliana.russano@inaf.it}
              \label{INAF-OAC}
        \and
        INAF - Astrophysical Observatory of Catania, Catania, Italy
        \label{INAF-OACt}
        \and
        Institute of Physics, University of Graz, Graz, Austria
        \label{UniGRAZ}
        \and
        INAF - Astrophysical Observatory of Torino, Pino Torinese (TO), Italy
        \label{INAF-OATo}
        \and
        INAF - Astronomical Observatory of Trieste, Trieste, Italy
        \label{INAF-OATs}
        \and
        Universidad de Mendoza, CONICET, Grupo de Estudios en Heliofisica de Mendoza, Mendoza, Argentina
        \label{CONICET}
        \and
        University of Florence, Sesto Fiorentino (FI), Italy
        \label{UniFi} 
         \and
        INAF – Astronomical Observatory of Arcetri, Florence, Italy
        \label{INAF_Arcetri}
        \and
        INAF – Istituto di Astrofisica Spaziale e Fisica Cosmica, Milan, Italy
        \label{INAF-IASF}
             }

  \abstract
   {The study of the kinematic and dynamic evolution of fast, eruptive events from the middle to the high solar corona is one of the primary scientific objectives of the Metis coronagraph on board the Solar Orbiter satellite. During the spacecraft's perihelion passages, Metis operates in an observational mode that acquires visible light images at a cadence of 20\,seconds, achieving a spatial resolution of around 2000\,km at a solar distance of 0.28\,au. This capability enables the detailed capture of coronal structures and transient events, such as coronal mass ejections (CMEs) with unprecedented spatial and temporal resolution.}
   {Between October 8 and 9, 2022, an extensive CME was observed in the Metis plane of the sky while Solar Orbiter was at a distance of 0.3\,au, allowing for a spatial resolution in the visible channel of 4.4$\cdot 10^3$\,km/pixel. We aim to exploit the unprecedented high-resolution of Metis observations to resolve multiple substructures within the CME front, revealing plasma elements propagating at different speeds and along distinct trajectories, thus enabling a detailed kinematic characterisation of the eruption.  
   }
   {To highlight the complex morphology of the solar eruption observed in Metis images, a  normalization-based running difference enhancement algorithm was applied. Height-time diagrams have been implemented to estimate the propagation speeds and frequency variations of newly emerging features. In addition, a three-dimensional reconstruction model of the flux rope structure, combined with data from other space-based coronagraphs and disk imagers, enabled tracking of the entire CME evolution from its early phase in the lower corona up to the middle corona (approximately 5 solar radii). Joint observations with Solar Orbiter/EUI-FSI provided insights into the eruption's initiation in the inner corona, while Metis' high-resolution imaging captured its development into the middle corona, allowing a comprehensive view of CME kinematics across multiple coronal layers. 
   }
   {The high spatial and temporal resolution observations provided by Metis make it possible to study the fine structure of the CME, highlighting the internal motions of the coronal plasma and characterizing its kinematic evolution. The results obtained from this study  contribute to a deeper understanding of the morphology and kinematics of CMEs. 
   Furthermore, the detection of circular, fast-propagating wavefronts (travelling at $\sim$500\,km/s with a characteristic period of $\sim$3 minutes) at the western flank of the CME front opens new interpretative scenarios, suggesting the involvement of wave excitation and magnetic field reconfiguration processes in shaping the CME evolution.
   Multiple interpretations are proposed for the observed coronal plasma wave-trains, including the presence of quasi-periodic propagating fast modes, providing new insights into wave generation and energy transport in the solar corona.
   }
  {}


   \maketitle

\section{Introduction}
\label{sec:intro}

Coronal mass ejections (CMEs) are defined as large-scale expulsions of magnetized plasma from the lower corona into the interplanetary medium. 
Since the advent of space-based observations, they are among the most extensively studied solar eruptive phenomena \citep{Howard_CME_history_2023FrASS..1064226H}.
Historically, our understanding of the large-scale  morphology and structure of CMEs comes from space-based white-light coronagraphs, such as STEREO/SECCHI COR1 and COR2 \citep{STEREO_2008SSRv..136....5K,SECCHI_2008SSRv..136...67H} and SOHO/LASCO C1, C2 and C3 \citep{SOHO_1995SoPh..162....1D,LASCO_1995SoPh..162..357B}, which offer complementary vantage points and continuous solar monitoring. 
These instruments enabled extensive statistical studies of CME kinematics from the standpoint at 1\,au and laid the foundation for the current physical comprehension and theoretical modelling of these transient phenomena evolution. 

The observed appearance of CMEs in white-light images results from Thomson scattering effects, which describe the interaction of sunlight with free electrons present in the corona and CME plasma. 
The geometry of the CME relative to the Sun-observer line (which affects projection), as well as the polarisation properties of the scattered light, further modulate its appearance, enhancing or suppressing specific CME features \citep{Vourlidas_CME_bright_2006ApJ...642.1216V,Thomson_surface_2012ApJ...752..130H}. This results in a wide range of observed morphologies, with brightness variations expected to be substantially influenced by the CME's location relative to the Thomson surface (i.e. the sphere of maximum scattering efficiency, \citealp{Vourlidas_CME_bright_2006ApJ...642.1216V}).
 
The standard CME structure consists of three distinct components: a frontal loop, an inner dark cavity, and a dense bright core, which is often aligned with an erupting prominence \citep{Chen_CME_review_2011LRSP....8....1C,Webb_CME_review_2012LRSP....9....3W}. 
Moreover, the appearance of a CME is strongly influenced by various geometrical factors mainly depending on their magnetic configuration and projection effects \citep{Forbes_2000JGR...10523153F,cremades_3D_CME_2004A&A...422..307C}. 
Certain CMEs, exhibit more intricate geometries, possibly depending on their initiation mechanisms and surrounding ambient corona. Some CMEs appear as narrow jets, while others, known as streamer blowout CMEs, arise from the disruption of pre-existing coronal streamers \citep{Vourlidas_SBO_2018ApJ...861..103V}. Halo CMEs, observed as large-scale and outward expansions, arise when the eruptions are directed along the Sun-observer line, causing significant projection effects.

A substantial amount of research has focused on exploring the internal structure of these phenomena from a magnetic field perspective and interaction with the surrounding coronal environment. 
Understanding their formation mechanisms and evolution requires detailed studies of their three-dimensional (3D) structure and the origin, as the role of magnetic flux ropes and prominences at the core \citep{cremades_3D_CME_2004A&A...422..307C,Vourlidas_CME_flux_rope_2013SoPh..284..179V,Howard_CME_shape_on_POS_2017ApJ...834...86H}. 

With the advent of high-resolution imaging instruments operating at closer distances from the Sun, such as those onboard Solar Orbiter (SolO, \citealp{SolO_2020A&A...642A...1M}) and  Parker Solar Probe \citep{PSP_2016SSRv..204....7F}, CME observations have undergone a paradigm shift \citep{Howard_PSP_FR_2022ApJ...936...43H}. 
The shift is characterized by moving from traditional, 1\,au vantage viewpoint coronagraphy to multi-perspective and close-proximity observations that capture CMEs in their early acceleration and formation phases. The improved spatial and temporal resolution of these instruments enables direct observation of fine-scale structures \citep{Cappello_2024A&A...688A.162C}, magnetic reconnection signatures, and rapid dynamical evolution, previously unresolved in SOHO or STEREO-era data.

Among the key instruments onboard SolO, the Metis coronagraph \citep{metis_instr} has been acquiring images of the solar corona in two wavelength bands since 2020. This externally occulted coronagraph is equipped with two imaging channels: a visible light (VL) channel, sensitive to Thomson scattering of photospheric light by coronal electrons (broadband 580-640\,nm), and an ultraviolet (UV) channel, which detects \ion{H}{i}~\Lya\ line emission at 121\,nm, tracing resonantly scattered radiation from neutral hydrogen atoms.  
When operating at heliocentric distances between 0.28 (resulting in a field of view (FoV) ranging from 1.7 to 3.6 solar radii (R$_{\sun}$)) and 0.5\,au (with related FoV from 3.0 to 6.4\,R$_{\sun}$), Metis provides up to 20 times higher spatial resolution than that achievable at 1\,au. 

\begin{figure*}
   \centering
   \includegraphics[clip, trim=0.1cm 5cm 1cm 5cm,width=18cm]{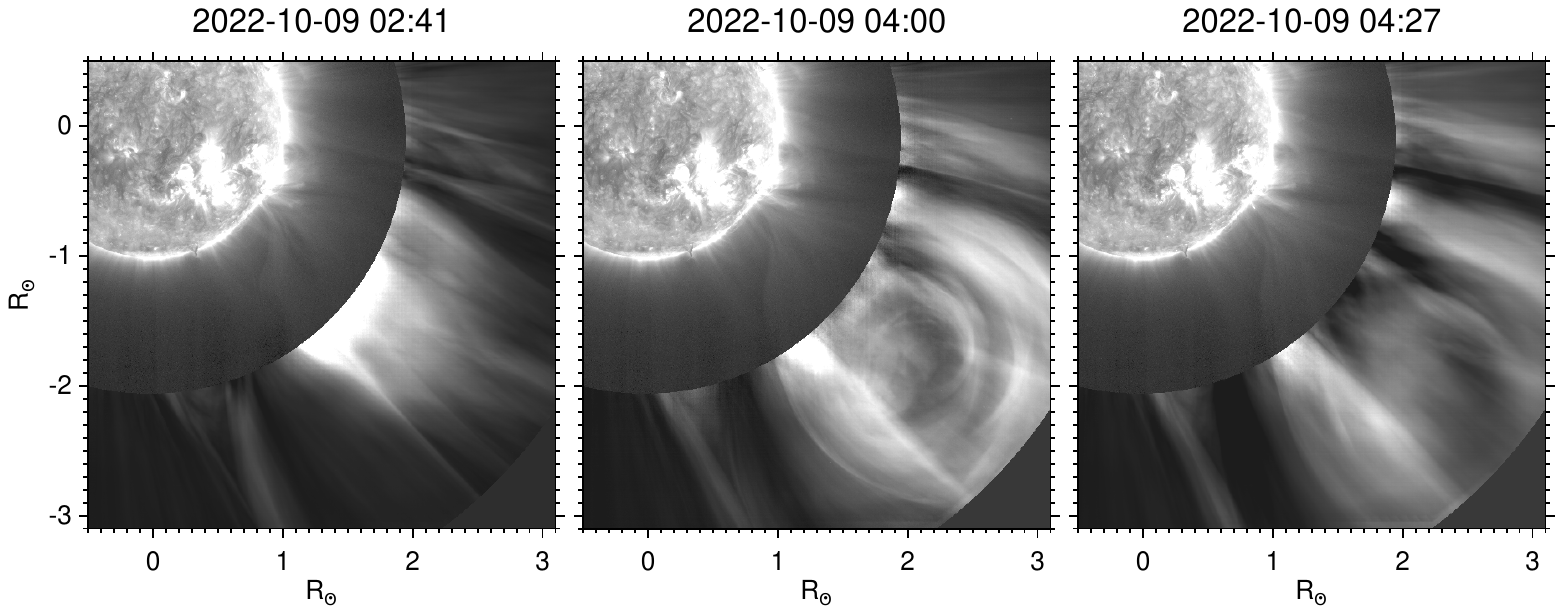}
   \includegraphics[clip, trim=0.1cm 3.5cm 1cm 4cm,width=13cm]{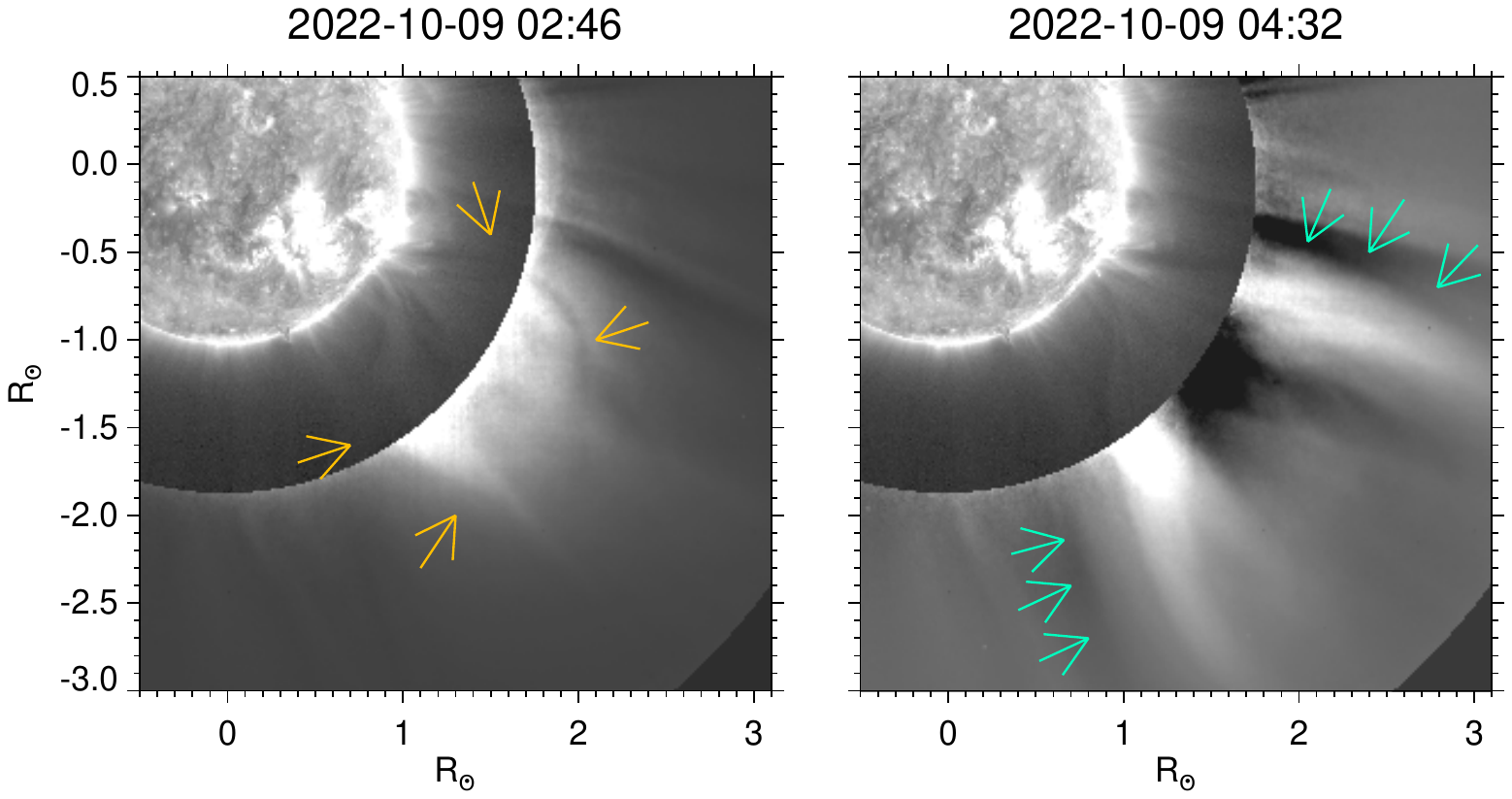}
   \caption{Composite images of the disk observed by SolO/EUI-FSI at 174\,$\r{A}$ and the corona observed by Metis during the event on 9 October 2022. The first row displays Metis VL channel images in total brightness (the associated movie is available online Metis\_tB\_EUI\_FSI174.mp4), while the second row shows the UV channel images. Orange arrows indicate the UV dark ring edges. Green arrows denote the flanks of CME as seen in the UV FoV.
   SolO/EUI-FSI images have been processed using the Multiscale Gaussian Normalization algorithm (MGN, \cite{MGN}). Metis total brightness images have been normalised after subtracting a minimum background image created for this set of images. Metis UV images are shown as base difference images, where the base frame used to subtract the contribution of the K-corona was acquired on 8 October 2022 at 18:10\,UT.
   The SolO/EUI-FSI data used in this work are taken from \cite{euidatarelease6}.}
              \label{fig:EUI_Metis_tB}%
    \end{figure*}

The present study aims to demonstrate the unique capability of Metis to image the middle corona at high spatial and temporal resolution, revealing fine-scale structures within CMEs that have remained unseen by previous coronagraphs. Leveraging its specific high-resolution white-light observation modes, Metis enables us to resolve both the morphology and kinematics of CMEs with unprecedented detail, including the identification of substructures within the magnetic flux rope and the surrounding plasma environment. 
A particularly remarkable achievement highlighted in the present work is the Metis’ ability to capture coronal wave-trains propagating laterally to the CME front. 
Wave-like disturbances, closely associated with solar eruptive events, provide crucial diagnostics for understanding the CME dynamics, energetics, shock formation and heating processes of the outermost solar atmosphere \citep{Warmuth_waves_2015LRSP...12....3W,Jess_waves_review_2023LRSP...20....1J}. Previous research using extreme ultraviolet (EUV) imagers such as SOHO/EIT \citep{EIT_1995SoPh..162..291D}, STEREO/SECCHI EUVI-A \citep{EUVI_2008SSRv..136...67H}, and SDO/AIA \citep{SDO_2012SoPh..275....3P} has allowed for the tracking of EUV wave propagation across the solar disk. Unlike EUV imagers, which capture chromospheric and lower coronal wave signatures, as will be demonstrated below, Metis allows direct imaging of coronal wave-trains in white-light, providing insight into their propagation in the middle corona and their potential role in magnetosonic wave modes generation.

Recent works by the Metis team have also explored the potential of high cadence observations not only in the early evolution of the CME \citep{Bemporad_CME_HR_2025} but also for the first time detection of wave-like density fluctuations in coronal structures like streamers and pseudo-streamers \citep{Andretta_waves_2025}.

In Sec.\,\ref{sec:observation_desc} we present the description and the morphology of the CME under study. 
Sec.\,\ref{sec:HR_res} is dedicated to the study of the kinematics of the event and its details, as evidenced by the Metis VL high-cadence observations.
In Sec.\,\ref{sec:results}, a comprehensive synthesis of observations is presented, along with the morphological analysis of the event and the detailed interpretation of the coronal wave-trains. 
A summary of the main findings is outlined in Sec.\,\ref{sec:conclusions}.

\begin{table*}
\caption{Observing parameters for the event as seen by Metis.}             
\label{table:event_info}      
\centering          
\begin{tabular}{c c | c c | c c | c c | c c | c c | c c}   %
\hline\hline       
Event & Time & \multicolumn{2}{c}{Binning} & \multicolumn{2}{c}{DIT} & \multicolumn{2}{c}{NDIT} &  \multicolumn{2}{c}{T$_{exp}$} & \multicolumn{2}{c}{Cadence} & \multicolumn{2}{c}{Spatial scale}\\ 
date & [UT] & & & \multicolumn{2}{c}{[s]} & & & \multicolumn{2}{c}{[min]} & \multicolumn{2}{c}{[s]} & \multicolumn{2}{c}{[10$^3$ km/pxl]} \\
\cline{3-14}
  9 Oct. 2022 & 00:17 - 12:00 & VL & UV & VL & UV & VL & UV & VL & UV & VL & UV & VL & UV \\
\hline                    
 \multicolumn{2}{c}{WIND Obs. mode}  & 2x2 & 2x2 & 30 & 60 & 7 & 7 & 3.5 & 7 & 960 & 480 & 4.5 & 9 \\  
 \multicolumn{2}{c}{FLUCTS - TBF Obs. mode} & 2x2 & - & 20 & - & 1 & - & 0.33 & - & 20 & - & 4.5 & - \\
\hline                  
\end{tabular}
\tablefoot{The table provides: the event date and time range in the Metis FoV. The data acquisition settings (binning, detector integration time (DIT), the  number of images averaged onboard (NDIT), exposure time (T$_{exp}$), cadence, and spatial scale of the images after binning for both Metis channels and observation modes) are on the right side of the table. WIND and FLUCTS - TBF observation modes are described at the beginning of Sec.\ref{sec:observation_desc}.}
\end{table*}

\section{Observational study of the 8–9 October 2022 coronal mass ejection}
\label{sec:observation_desc}

Between 8 and 9 October 2022, the CME under investigation erupted from the southwestern limb of the Sun and was observed in succession by multiple instrument. SolO/EUI-FSI \citep{EUI_2020A&A...642A...8R} detected the event around 22:00\,UT on 8 October; Metis observed the CME entering its field of view from the southwest by 00:17\,UT on 9 October, tracking it until approximately 12:00\,UT, after which only residual plasma flow was detected. The CME remained visible until 18:00\,UT the following day, when it reached the end of the field of view of SOHO/LASCO-C3, the coronagraph with the widest coverage (up to 30\,R$_\odot$). 
Observations from Metis on Solar Orbiter, positioned at 0.3\,au, captured the event within an annular field of view spanning 1.8\,-\,3.9\,R$_\odot$.

During this period, Metis operated in a dedicated observational mode designed to track coronal wind flows, both slow and fast streams (namely WIND observation mode, \citealp{metis_instr}) \footnote{\href{https://www.cosmos.esa.int/web/solar-orbiter/soops-summary}{The Solar Orbiter Observing Plan} (SOOP) for the 8 and 9 October 2022, performed in coincidence with other SolO instuments, was L\_FULL\_HRES\_HCAD\_Coronal-Dynamics. This SOOP focuses on the observation of structures in the outer corona and linking them to the in-situ heliosphere observations.}.
Images were acquired at 16\,min cadence in the VL channel and 8\,min cadence in the UV channel, corresponding to exposure times of 3.5\,min and 7\,min, respectively, over a 12-hour observation period.
This observational sequence was briefly interrupted for a  40-minute high-resolution total brightness acquisition (namely FLUCTS-TBF mode, \citealp{metis_instr}) from 03:30\,UT to 04:10\,UT, during which time the UV channel was inactive. 
VL total brightness images were produced using four quadruplets of linearly polarized images, calibrated following the in-flight radiometric calibration procedure in \cite{DeLeo_VL_2023A&A...676A..45D}. 
During this high-resolution mode, images were acquired every 20 seconds, optimizing the investigation of density fluctuations in the middle corona (see also \citealp{Andretta_waves_2025}). 
The combination of high temporal cadence and a pixel-scale of 4.4\,$\cdot 10^3$\,km/pxl (after rebinning) allowed for an in-depth study of the CME's internal structure.
For each observational mode and spectral channel, individual images were generated through an on-board averaging process of NDIT images, captured with designated Detector Integration Times (DITs) to ensure a total exposure time matching the total cadence (see Tab.\,\ref{table:event_info}).
UV channel images were instead calibrated according to \cite{DeLeo_UV_2025A&A...697A..73D}.

Figure\,\ref{fig:EUI_Metis_tB} shows the evolution of the event, displaying the composite images of SolO/EUI-FSI observations at 174\,$\r{A}$ and Metis total brightness (top panels) and UV images (bottom panels).  
The first and third frames in the top row, acquired in the standard WIND observation mode, exhibit slightly blurred structures, due to the longest integration time, compared to the middle frame, which was obtained during the FLUCTS-TBF mode with only 0.33-minute integration time. 
The central image distinctly reveals well-defined structures within the evolving CME, that will be discussed in detail in Sec.\,\ref{sec:HR_res}. 
These three frames are base differences normalised using an image \( B_{\text{norm}}(r) \) derived by averaging the values of the minimum background frame \( B_m \) along the azimuthal direction, where \( r \) represents the distance from the center of the solar disk. 
The background image \( B_m \) was derived by calculating the pixel-by-pixel first percentile across the brightness image series. 

Two movies of the Metis total brightness images are available as supplementary material: Metis\_tB\_HR.mp4 shows only the high resolution 40-min sequence enhanced as described in Sec.\,\ref{sec:HR_res}; Metis\_tB\_EUI\_FSI174.mp4 shows the whole sequence from which the images in the first row of Fig.\,\ref{fig:EUI_Metis_tB} are taken (the associated movies are available online).

In both Metis VL and SolO/EUI-FSI images, the eruption appears as a typical CME structure. A nearly spherical envelope with sharp contours relative to the surrounding plasma undergoes inflation from the solar surface in the SolO/EUI-FSI images, before continuing its expansion within the Metis field of view. 
The CME front presents a well-defined, circular structure in the Metis field of view, characterized by a void in connection with a highly structured core. Helical field lines embedded within the bright leading edge indicate a magnetic flux rope structure.
The maximum angular width of the CME in the plane of sky (PoS) is $\sim$53 degrees.

\begin{figure}
   \centering
   \includegraphics[clip, trim=5cm 9cm 5cm 9cm,width=5cm]{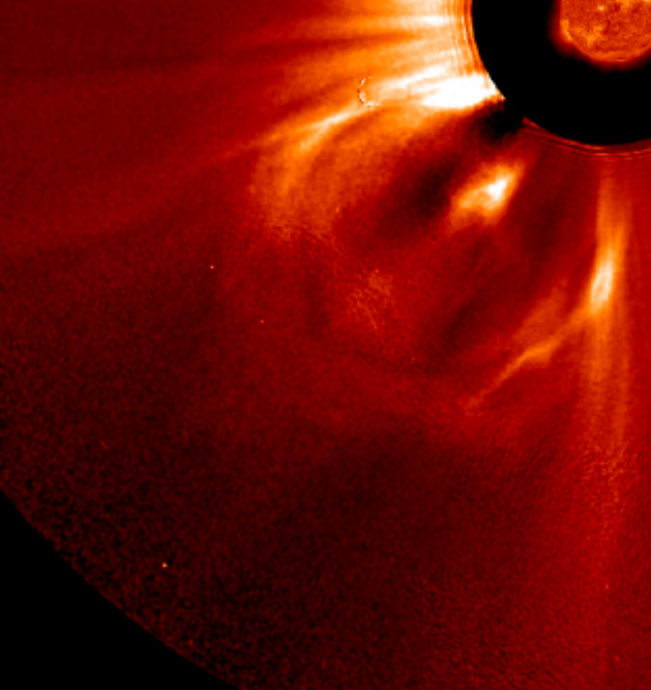}
   \caption{Coronal mass ejection observed by STEREO/EUVI disk imager and COR2 coronagraph at 7:23\,UT on 9 October 2022. The image is produced with the open-source software JHelioviewer \citep{JHV}.}
              \label{fig:COR2}%
    \end{figure}

The core of the CME does not appear particularly bright in Metis UV channel (second row of Fig.\,\ref{fig:EUI_Metis_tB}), suggesting that no cool prominence eruption is present. 
In contrast, the CME exhibits a classic three-part structure in STEREO/SECCHI COR2-A white light coronagraph field of view (see Fig.\,\ref{fig:COR2}), where a well-defined bright core and a large dark cavity are visible, starting about an hour after the structure left the FoV of Metis. 
This evidence will be further interpreted in Sec.\,\ref{sec:morphology} in light of the morphological characteristics of the event.

On the other hand, a dark ring, aligned with the eruption envelope, is clearly visible in the Metis UV channel image (left panel of the second row in Fig.\,\ref{fig:EUI_Metis_tB}, indicated by orange arrows). This alignment is further accentuated in the adjacent SolO/EUI-FSI 174\,$\r{A}$ FoV. The thickness of the ring is $\sim$\,0.06\,R$_\odot$ ($\sim$\,45\,Mm).
The structured plasma flow (speed around 250\,km/s as shown in Sec.\,\ref{subsec:velocity}) is also visible in the remaining UV images and in the associated movie available online \footnote{Metis UV images, available for further examination in the online movie Metis\_UV\_EUI\_FSI174.mp4, were enhanced with the Wavelets Optimized Whitening filter \citep{WOW_filter_2023A&A...670A..66A}.}, as the central part of the CME exits the field of view. 
There is a large gap of 40 min in UV frames that results from the high-cadence acquisition of the VL sequence.

\subsection{From the source region to the middle corona detection}
\label{sec:GCS}

\begin{figure}
   \centering
   \includegraphics[clip, trim=2.5cm 3cm 0.3cm 3cm,width=6cm,angle=270]{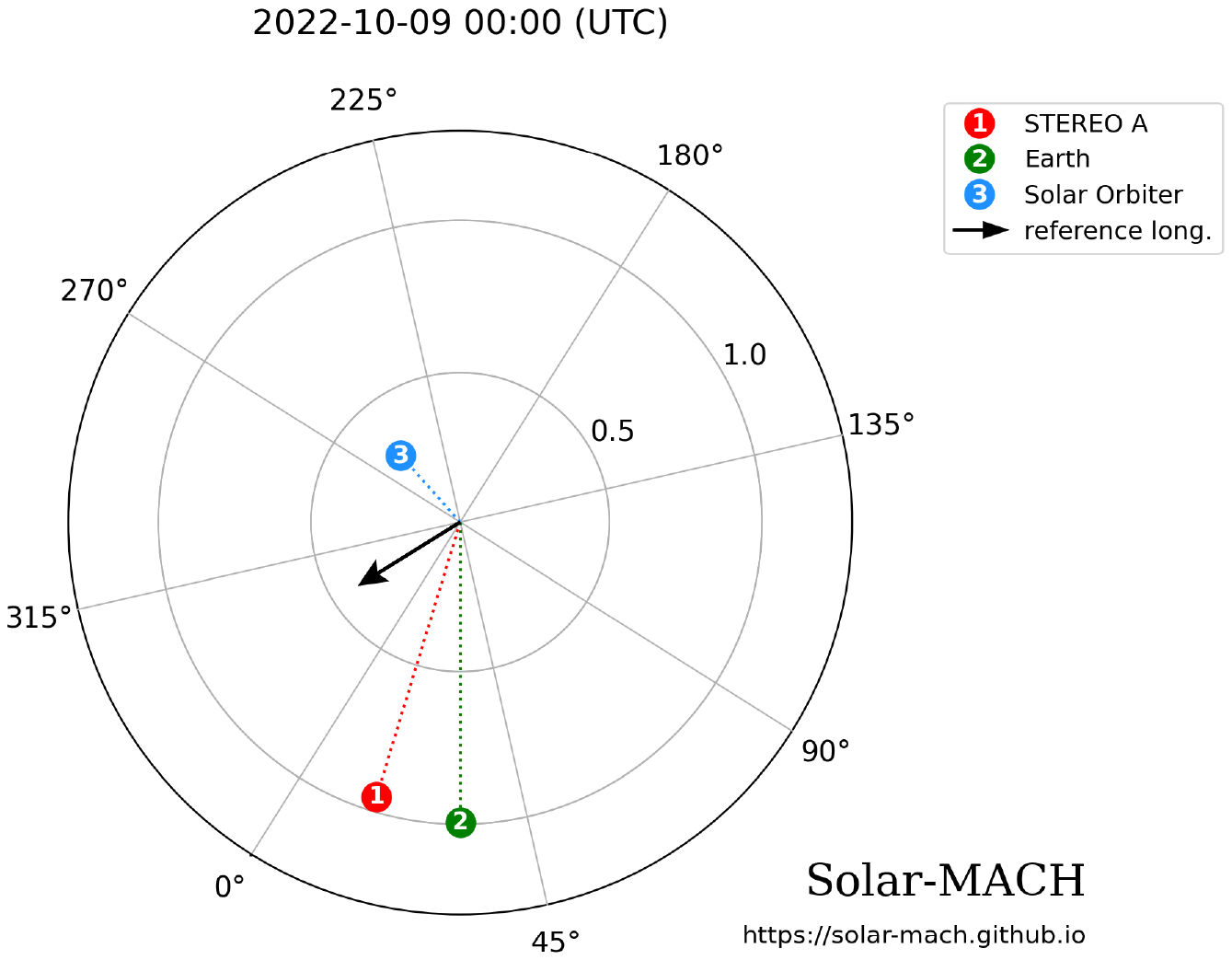}
\caption{Multi-spacecraft longitudinal configuration plot (Carrington coordinate System) for 9 October 2022, as provided by the Solar Mach web application \citep{SolarMach_2023FrASS...958810G}. The reference longitude (black arrow) represents the CME propagation direction (see Sec.\,\ref{sec:GCS} and Appendix\,\ref{appendix:GCS}).}
              \label{fig:orbit_plot}%
    \end{figure}

The relative orbital positions of other spacecraft with respect to Metis allow us to study the structure and kinematics of the event from different points of view (see Fig.\,\ref{fig:orbit_plot}). By combining coronagraphic VL images acquired by SOHO and STEREO-A, together with EUV disk imagers STEREO/SECCHI EUVI-A, SDO/AIA, SDO/HMI \citep{SDO_HMI_2012SoPh..275..207S} and GOES-R (GOES-16) Solar Ultraviolet Imagers (SUVI, \citealp{SUVI_2022SpWea..2003044D}), it is possible to infer the source region of the event and follow its evolution in the low corona, as well as the magnetic configuration underlying the eruption.

\begin{figure}
   \centering
   \includegraphics[clip, trim=3.8cm 1.8cm 3.8cm 2cm,width=5.1cm,angle=270]{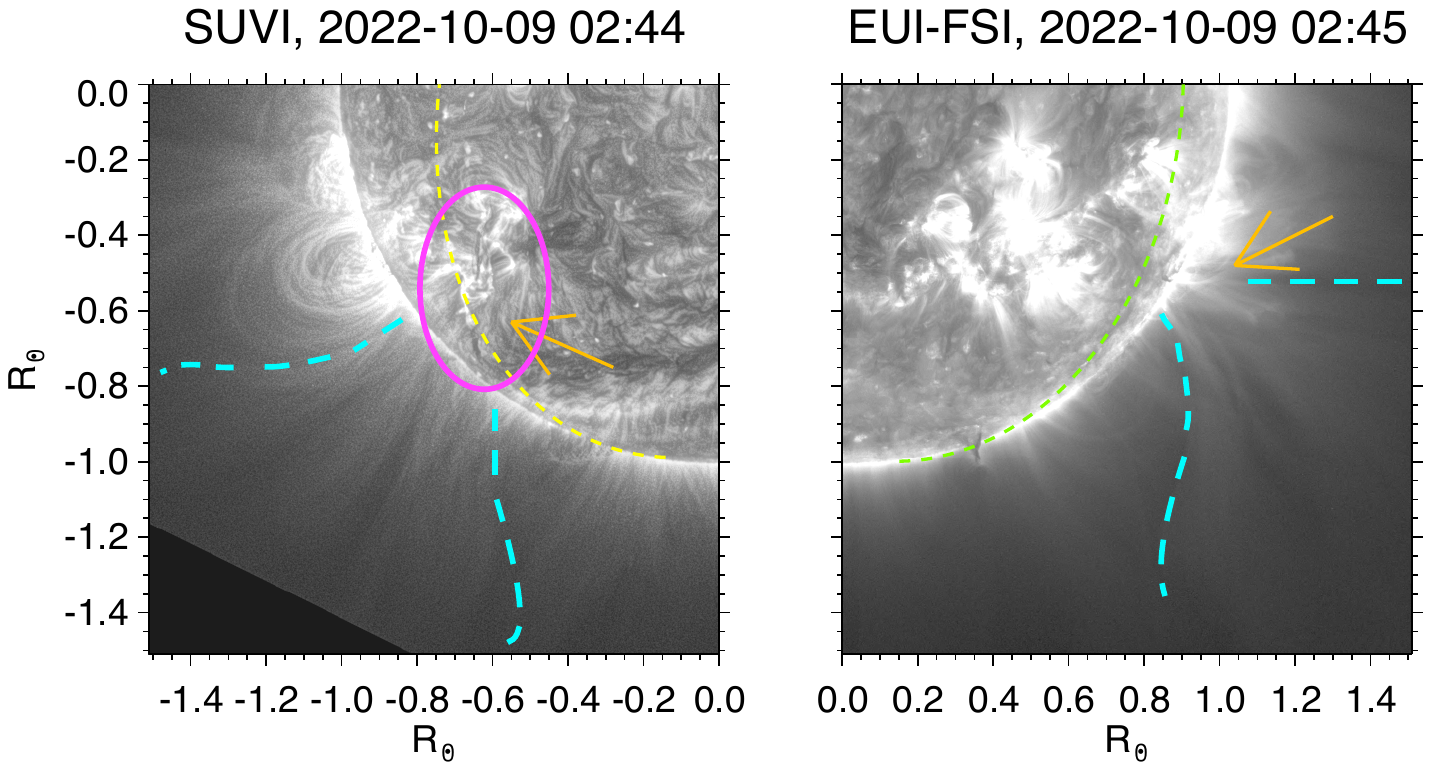}
   \caption{Evolution of the event as seen by GOES-R/SUVI (195\,$\r{A}$, the associated movie SUVI\_195.mp4 is available online) and SolO/EUI-FSI (174\,$\r{A}$) on the disk and in low corona around 2:44\,UT. Orange arrows point to the eruption source region. The magenta circle encloses the active region of interest.
   The yellow line in the left image indicates the solar limb as seen by Solar Orbiter. The green line in the right image indicates the solar limb as seen by Earth. The dark boundary of the CME envelope is highlighted by manually drawn cyan contours as it expands through the low corona. 
   Images are enhanced using the MGN algorithm. 
   }
              \label{fig:SUVI}%
    \end{figure}

\begin{figure}
   \centering  
   \includegraphics[clip,width=3.05cm,trim=12cm 5cm 0cm 14.3cm,angle=270]{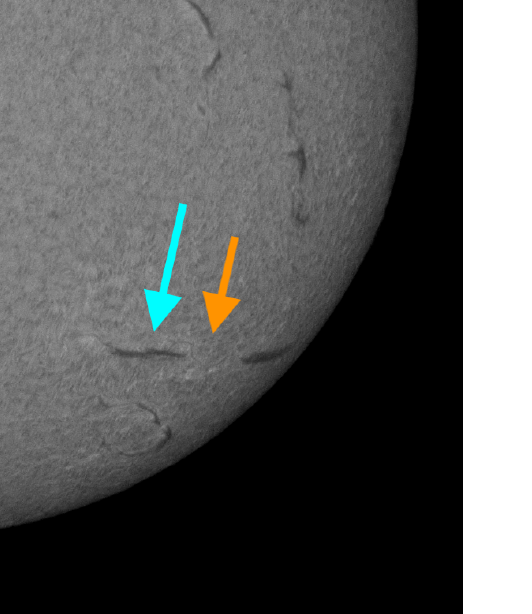}
   \includegraphics[clip,width=3.05cm,trim=12cm 5cm 0cm 14.3cm,angle=270]{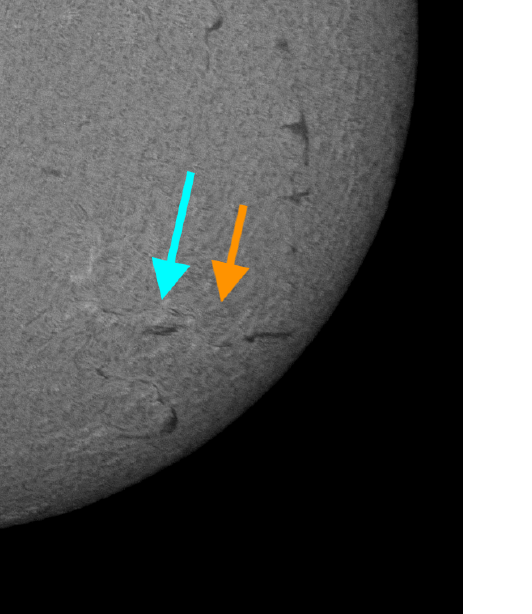}\hfill
   \includegraphics[clip,width=3.05cm,trim=12cm 5cm 0cm 14.3cm,angle=270]{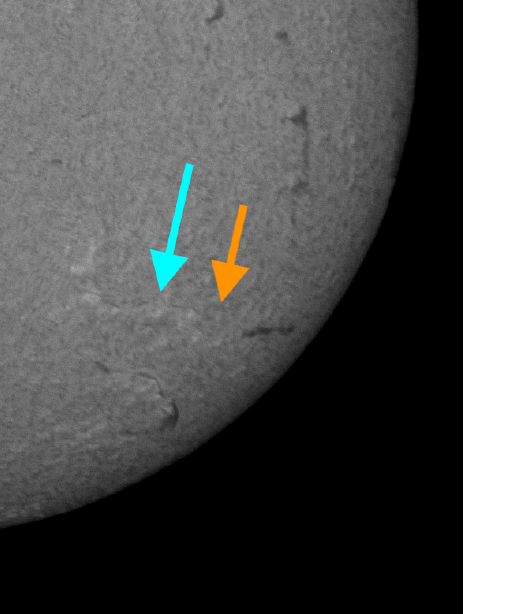}
   \includegraphics[clip,width=3.07cm,trim=6.5cm 6cm 3cm 10.3cm,angle=270]{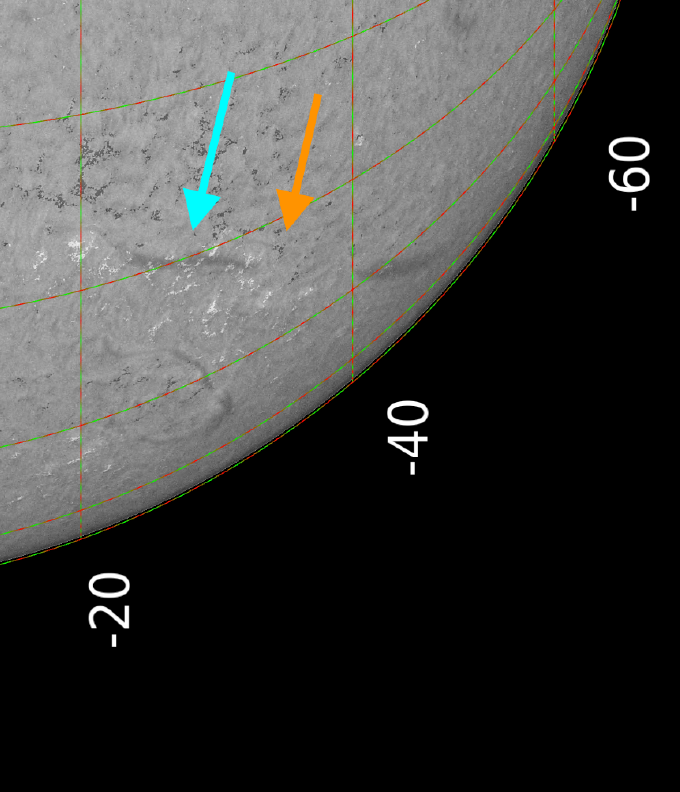}\hfill
   \includegraphics[clip,width=3.25cm,trim=6.5cm 6.5cm 2.5cm 10.3cm,angle=270]{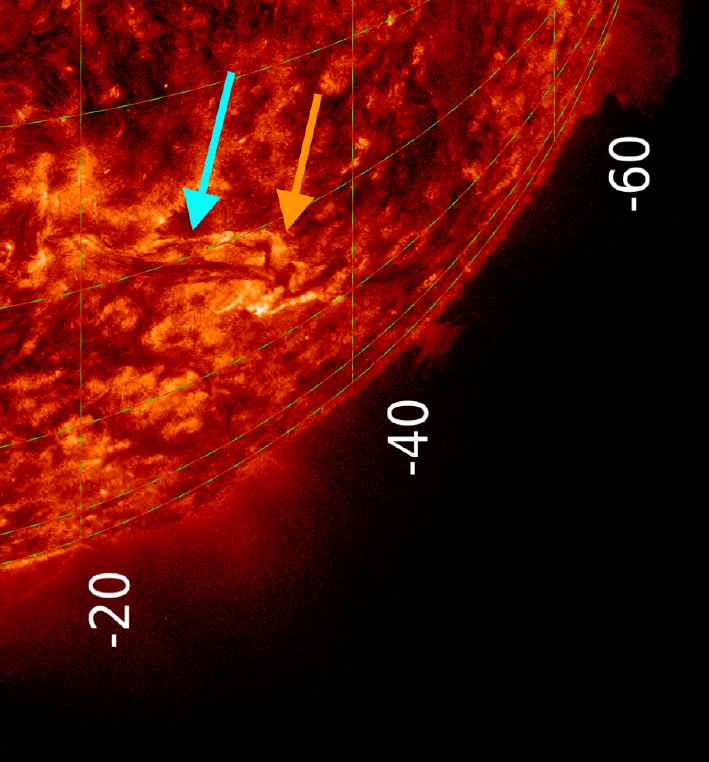}
   \includegraphics[clip,width=3.48cm,trim=4cm 5cm 0cm 7.3cm,angle=270]{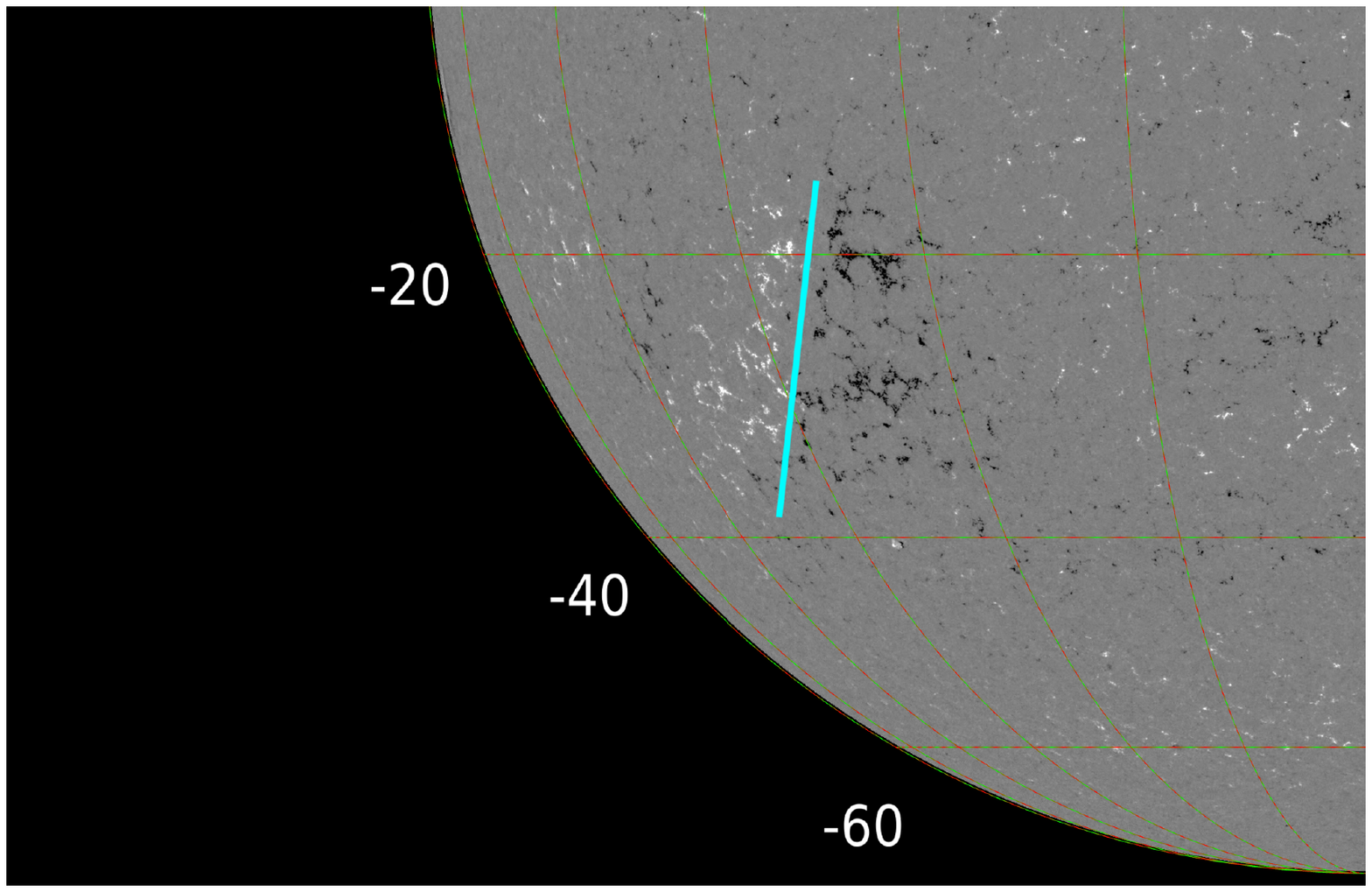}
   \caption{Zoomed images of the solar disk in the source region on 9 October 2022. 
   Top row: H-$\alpha$ (6562.8\,$\r{A}$) images from the NSO/GONG archive, recorded at Learmonth Solar Observatory in Australia (03:00\,UT, left) and Cerro Tololo Observatory in Chile (12:00\,UT, right).
   Middle row: Cerro Tololo H-$\alpha$ image at 14:00\,UT (left panel) and the Learmonth image of the first panel (03:00\,UT) on which the magnetogram measured by SDO/HMI was superimposed (right panel). 
   Bottom row: on the left, the source region as detected by SDO/AIA 304\,$\r{A}$; on the right, the SDO/HMI magnetogram, with the PIL approximated by a cyan line. 
   The last three images were created with the open-source JHelioviewer software. 
   The white numbers indicate the negative solar latitude. 
   The filament of interest is indicated by the cyan arrows. The source region is indicated by the orange arrows. These correspond to the arrow in the SUVI image in Fig.\,\ref{fig:SUVI}.}
              \label{fig:GONG}%
    \end{figure}

The 3D geometry of the eruption is characterized using the Graduated Cylindrical Shell (GCS) model \citep{Thernisien_2006ApJ...652..763T,GCS_implementation_2011ApJS..194...33T}. By combining data from multiple vantage points the model provides an empirical fit of the CME magnetic flux rope structure to recover the main geometrical parameters (details of the calculations are provided in Appendix\,\ref{appendix:GCS}).
The parameters relevant for our analysis are: the position of the CME flux rope apex at longitude 339$^\circ$ and latitude -32$^\circ$ in the Carrington Coordinate System, at a height of 4.6\,R$_\odot$.
The Carrington longitude is used in Fig.\,\ref{fig:orbit_plot} to indicate the CME's propagation direction (black arrow). This direction is closely aligned with the Metis PoS, implying that projection effects on the measured PoS velocities are minimal (see Sec.\,\ref{subsec:velocity}). 
The resulting GCS parameters are also in agreement with the approximate location of the source region, as demonstrated by the red crosses in the first panel of Fig.\,\ref{fig:GCS}, which correspond to the region indicated by the orange arrows in Figures \ref{fig:SUVI} and \ref{fig:GONG}.

The eruption source region is clearly visible in the south-eastern quadrant of GOES-R/SUVI disk images in the 195\,$\r{A}$ band, which were acquired with a 4-minute cadence (orange arrow in Fig.\,\ref{fig:SUVI}). 
In the corresponding frame of SolO/EUI-FSI, the same region appears almost at the limb (Fig.\,\ref{fig:SUVI}, right panel).
Lower in the chromosphere, the H-$\alpha$ images of the solar disk during the initial phase of the eruption (first two rows of Fig.\,\ref{fig:GONG}) reveal the presence of two filaments (henceforth the northern and southern filaments), aligned with the region where the CME originates (indicated with orange arrows in Fig.\,\ref{fig:GONG}).

SDO/HMI magnetograms (Fig.\,\ref{fig:GONG}, bottom row, right panel) reveal the source area as a bipolar magnetic field region, where the proxy location of the polarity inversion line (PIL) is indicated as a cyan line. 
The PIL extends along the northern filament, enclosing the source region identified by Carrington coordinates. This is likewise visible in the SDO/AIA 304\,$\r{A}$ image in the bottom row of Fig.\,\ref{fig:GONG} (left panel), where a post-eruptive arcade (PEA) is seen to extend from the source region toward the northern part of the PIL (see also the discussion in Sec.\,\ref{sec:morphology}).

These observations suggest that the two H-$\alpha$ filaments and the CME source region are located along the same PIL and are thus presumably magnetically connected.
Therefore, the source region is located roughly at the southern end of the northern filament, as inferred from the post-eruptive arcade, and from projecting the Carrington coordinates onto the solar disk.
It is worth noting that given the deflection, rotation, and over-expansion that CMEs may experience during eruption, the GCS projection on the solar surface is typically rather a proxy than an exact match to the associated PIL's location, orientation, and extension.

The northern filament starts to show instability around 09:00\,UT (top right of Fig.\,\ref{fig:GONG}) and then disappears from the H-$\alpha$ images around 14:00\,UT (left panel in the middle row of Fig.\,\ref{fig:GONG}), notably several hours after the CME had traversed the coronagraphic FoVs. 
It should be noted that, over the time range of the eruption no flares were reported from or close to this active region, either at the time of the eruption or in the preceding hours \footnote{Solar flare events for the date of interest were searched on the web-list \href{https://www.lmsal.com/solarsoft/latest_events_archive.html}{SolarSoft Latest Events} and in the SolO/STIX flare event \href{https://datacenter.stix.i4ds.net/view/flares/list}{list}.}. 

Furthermore, in the 304\,$\r{A}$ band of SolO/EUI-FSI (shown in the associated movie available online\,\footnote{The online movie EUI\_FSI\_304.mp4 is produced with the open-source software JHelioviewer \citep{JHV}.}, on the west limb between -20 and -40 degree), plasma motion is observed suspended above the southwest limb for several hours, while material is seen falling back onto the surface. However, no clear evidence of a coronal prominence total or partial eruption is detected in these images and in other EUV band imagers, such as SDO/AIA and STEREO/SECCHI EUVI-A, likely owing to the plasma temperature being outside the filter's response range or to projection effects. 

In the GOES-R/SUVI image sequence processed with the MGN algorithm (see Fig.\,\ref{fig:SUVI}), the CME front becomes visible in the low corona starting at 23:00\,UT of 8 October 2022. The accompanying movie available online (SUVI\_195.mp4) offers clear evidence of the CME's darker boundary, revealing how the eruption envelope can be distinctly detected. This observation is consistent with the data acquired by the SolO/EUI-FSI, providing a coherent picture of the dynamic processes involved in the solar eruption, even in the lower corona.

\subsection{Morphological and magnetic structure of the coronal mass ejection}
\label{sec:morphology}

Based on the information gathered from the images and analysis presented in Sec.\,\ref{sec:GCS}, as well as previous studies such as \cite{cremades_3D_CME_2004A&A...422..307C}, \cite{Thernisien_2006ApJ...652..763T} and \cite{Vourlidas_CME_flux_rope_2013SoPh..284..179V}, it is therefore possible to illustrate the morphology of the event. 

At $\sim$03:00 UT on 9 October, the coronal loops above the identified source region exhibit expansion and upward motion. Shortly thereafter, a system of loops consistent with a PEA becomes visible, suggesting that magnetic reconnection is taking place in the low corona beneath the erupting flux rope \citep{post_eruption_loop_Tripathi2004A&A...422..337T}.
The onset of this process is clearly visible in EUV emission detected by GOES-R/SUVI (see the associated movie available online SUVI\_195.mp4), and verified in other observations available in the EUV band (i.e. SDO/AIA and STEREO/SECCHI EUVI-A). 
The position of both the filament and the PEA highlights the location of the magnetic PIL in the underlying region (see Fig.\,\ref{fig:GONG}).
The formation of PEAs is understood as a consequence of CMEs, likely initiated during the magnetic reconnection process occurring at the source site. PEAs are considered reliable tracers of CME source regions and are often used to study the photospheric magnetic field configuration and investigate possible CME initiations mechanisms \citep{cremades_3D_CME_2004A&A...422..307C}.

The position and orientation of the neutral line significantly influence the projected two-dimensional topology of CMEs in PoS white-light coronagraph images.
As defined in CME models, the principal axis of the CME appears to be aligned with a large-scale helical magnetic flux rope originating from the source region. In this configuration, the prominence constitutes the lower part of the magnetic system, while the neutral line, although aligned with the prominence axis, can sometimes defer significantly \citep{cremades_3D_CME_2004A&A...422..307C}. 
During the eruption, the flux rope may undergo significant deformation and rotation, depending on the complexity of the surrounding magnetic field and the dynamics of its evolution.
In our particular case, the neutral line
is not perpendicular to the solar limb, as observed in both the Metis and STEREO/SECCHI COR2-A fields of view. This geometry results in an axial CME observed along its axis of symmetry, with most of the core material concentrated along the line of sight (Fig.\,15 in \citealp{cremades_3D_CME_2004A&A...422..307C}).

The outward expanding flux rope is expected to appear dark when its orientation is perpendicular to the PoS \citep{Vourlidas_CME_flux_rope_2013SoPh..284..179V}. 
This explains why the CME observed by Metis presents a bright and well-defined structure, with a slightly pronounced void, while in STEREO/SECCHI COR2-A coronagraph images, taken few hours later, it appears darker and less structured. 

The viewing geometry in Metis observations further reduces the visibility of the bright core, making it virtually indistinguishable in high-cadence VL images, while it is clearly evident in STEREO/SECCHI COR2-A observations, consistent with the classic three-part structure of the CME (see Fig.\,\ref{fig:COR2}). 
One possible explanation for the absence of observed prominence material (whether partially or fully erupted) is that the structure may have undergone rotation with respect to the line of sight, thereby giving rise to a denser and brighter material in STEREO-A's field of view as a result of the expansion.
The different FoV and closer distance of Metis with respect to STEREO/SECCHI COR2-A and SOHO/LASCO-C2 leads to differences in the appearance of the observed structures.

At lower coronal heights, the eruption's properties of this event may follow the scenario proposed by \cite{Sun_magn_reconn_2015NatCo...6.7598S}, which provides direct evidence of magnetic reconnection in a 3D configuration. 
Prior to reconnection, two oppositely directed magnetic loops are anchored on either side of the filament, forming a pre-eruption cavity similar to what is observed in SolO/EUI-FSI at the beginning of the event (first image of Fig.\,\ref{fig:EUI_Metis_tB} and Fig.\,2 in \citealp{Sun_magn_reconn_2015NatCo...6.7598S}).
As the cavity rises, the underlying loops of opposite polarity gradually converge, eventually triggering magnetic reconnection. This process leads to the release of free magnetic energy, the production of post-eruptive arcades, and an enhanced emission due to the restructuring of the magnetic field. 

In conclusion, unlike the cases discussed in the aforementioned studies, the release of energy through reconnection was weak and gradual, insufficient to produce a detectable impulsive flare signature. However, PEAs can be observed, confirming that the CME originates from this region and is related to a flux rope system instability.

\section{Coronal mass ejection evolution with the high-cadence Metis observations}
\label{sec:HR_res}

In this section, we describe and analyze the sequence of high-cadence Metis observations, which enables us to characterise the kinematic evolution of the plasma both at the CME front and its surrounding regions, leveraging its high spatial and temporal resolution.

\begin{figure}
   \centering
   \includegraphics[clip, trim=4cm 1cm 3cm 0cm,width=9.5cm]{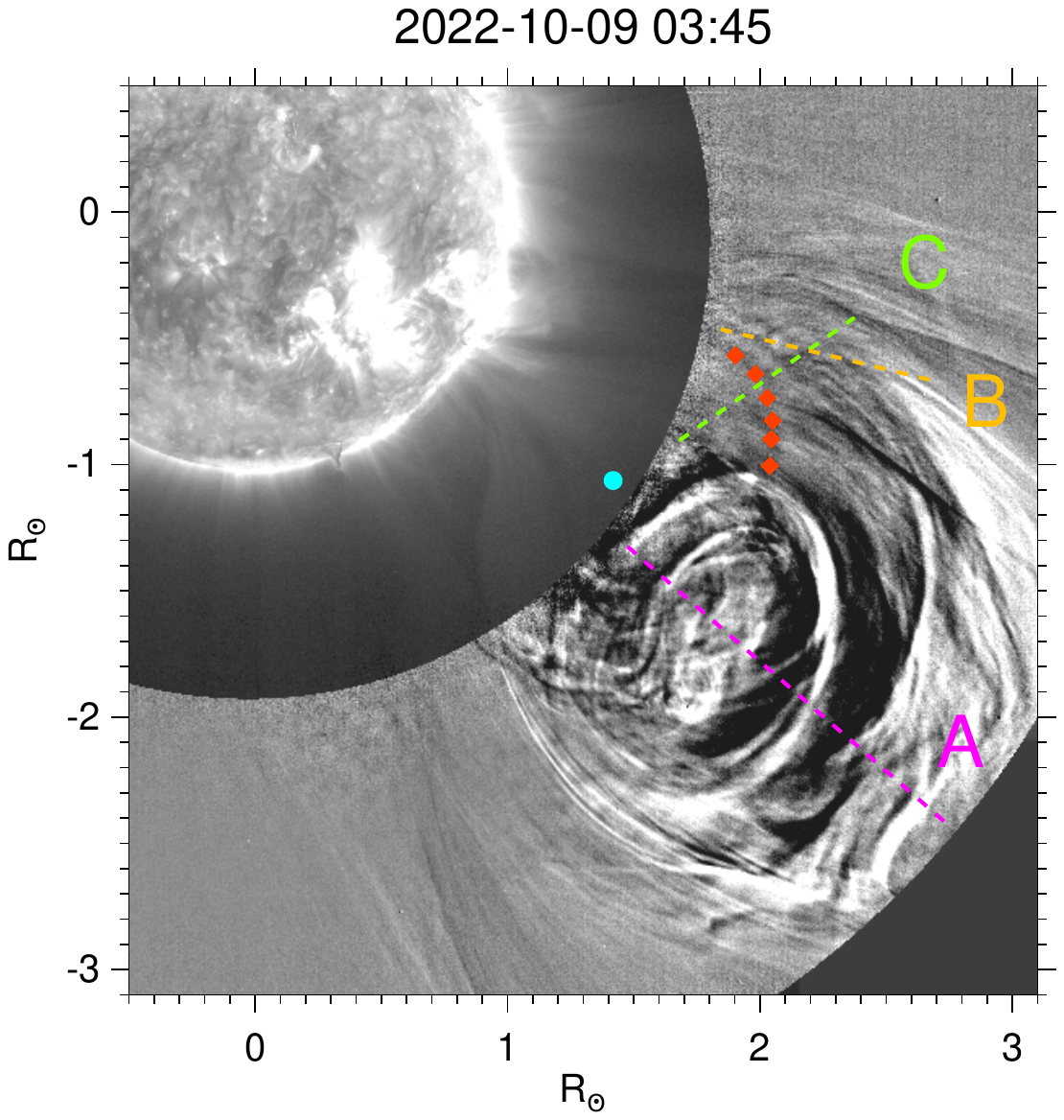}
   \caption{Single frame from the high-resolution Metis total brightness observation mode, acquired at 03:45\,UT, after application of the normalized running difference algorithm. The corresponding SolO/EUI-FSI 174 \,$\r{A}$ frame is overlaid. 
   Magenta line (feature A): image cut used to build the J-map of the CME flux rope along the central position angle in the top panel of Fig.\,\ref{fig:j-maps}. The radial distances of the edges of the segment are 1.7 and 3.2\,R$_\odot$  respectively.
   Orange line (feature B): image cut used to build the height-time diagram of the vortex-like plasma feature (bottom panel of Fig.\,\ref{fig:j-maps}). The radial distances of the edges of the segment are 1.9 and 2.8\,R$_\odot$ respectively.
   Green line (feature C): image cut used to build the J-map of circular wavefronts in Fig.\,\ref{fig:j-map_waves}. The radial distances of the edges of the segment are 1.9 and 2.4\,R$_\odot$ respectively.
   Red diamonds: mark a single circular wavefront. 
   Cyan dot: indicates the center of the circular wavefront estimated as explained in Sec.\,\ref{subsec:coronal_waves}. 
   To further appreciate the visible features, please refer to the associated movie available online Metis\_tB\_HR.mp4, where the high-resolution observation sequence is shown twice. In the right panel of the video, the segments of this figure are superimposed to help track the propagation of the observed structures.
   }
              \label{fig:Metis_HR}%
    \end{figure}

Figure\,\ref{fig:Metis_HR} presents a representative frame from the Metis total brightness sequence, acquired in FLUCTS-TBF observational mode, combined with the corresponding SolO/EUI-FSI 174\,$\r{A}$ frame. 
The Metis total brightness image has been processed using a normalised running difference algorithm, following a method similar to that described in detail in \cite{Andretta_waves_2025}. 
To enhance signal clarity and suppress noise, we applied a moving average over $2s+1$ frames to the time series of each pixel; we then computed a running average of the frames, taking care of subtracting to each image $k$ the previous independent images $k-(2s+1)$. The result is then normalised with the profile $B_\mathrm{norm}(r)$ created as described in Sec.\,\ref{sec:observation_desc}.   
In this analysis, \( s \) was varied between 3 and 5 depending on the signal-to-noise characteristic. 

This processing technique enabled the identification of numerous structures throughout the observational sequence, many of which had not been previously reported in similar events. 
The detailed evolution of plasma inside and in the neighbourhood of the main CME body is evident in the associated movie available online (Metis\_tB\_HR.mp4). 
We define three key kinematic structures associated to the CME, whose propagation path is highlighted by coloured dashed segment in Fig.\,\ref{fig:Metis_HR}. These segments are also used to build the height-time diagrams in Sec.\,\ref{subsec:velocity}. The three features will be referenced throughout this work in this way: feature A (magenta segment) identifies the central position angle of the CME flux rope and the accompanying compact and coherent plasma pile-up; feature B (orange segment) tracks rotational and downward motion of the plasma along the CME flank with a vortex-like morphology (this feature is further highlighted by orange arrows in Fig.\,\ref{fig:vortex} and in the associated movie available online Metis\_tB\_HR.mp4 right panel, while is discussed in Sec.\,\ref{subsec:KHi}); and feature C (green segment) delineates circular wave-trains propagating ahead of the CME front which are discussed in more detail in Sec.\,\ref{subsec:coronal_waves} and \ref{subsec:QPF}.
The waves detected in the running difference movies could be barely seen also in the base difference movies; however the running difference algorithm employed here is much more effective in enhancing these features. We also inspected other CMEs observed at high cadence, but we have not found similar wave-like features.

Finally, in the low corona part of the main circular front, faster plasma motions are observed, both on the sides of the front and in the centre. 
A second, more rapidly propagating front emerges at the base of the primary CME front around 03:58 UT. This secondary feature appears to interact with the rear of the primary CME flux rope before dissipating in the subsequent sequence of images at a reduced cadence (see the associated movie available online Metis\_tB\_EUI\_FSI174.mp4).
The onset of these movements begin approximately halfway through the observational sequence; however, the difficulty of visualisation and the absence of other observations in different instruments make their interpretation difficult to provide.

The nature of the periodic density perturbations visible in the eastern streamer in Fig.\,\ref{fig:Metis_HR} are unrelated to the CME evolution (they are present also in similar high-cadence observations taken on October 8 and 10), and have been investigated by \cite{Andretta_waves_2025} and will not be discussed in this work.

\subsection{Speed estimation}
\label{subsec:velocity}

This section presents the methodology and results for estimating the PoS speed of the structures highlighted in Fig.\,\ref{fig:Metis_HR} with capital letters A and B. For structure C, the method is applied with some differences that are discussed in Sec.\,\ref{subsec:coronal_waves}.
The velocity analysis was conducted using 'J-maps', where a radial segment was selected from a sample image based on the polar angle corresponding to the trajectory of the moving structure under study.  
Each segment was then extracted from every frame in the sequence (with a cadence of 20\,s) and  the intensity was averaged across the width of the segment to construct the J-maps (see dashed segments in Fig.\,\ref{fig:Metis_HR}). 
Finally, the speed was derived by applying a second-order polynomial fit to eight key points manually selected along the feature's trajectory.
The estimated speed values are summarised in Tab.\,\ref{tab:velocity}.

\begin{figure}
   \centering
   \includegraphics[clip, trim=4cm 2cm 3cm 3.cm,width=9.1cm]{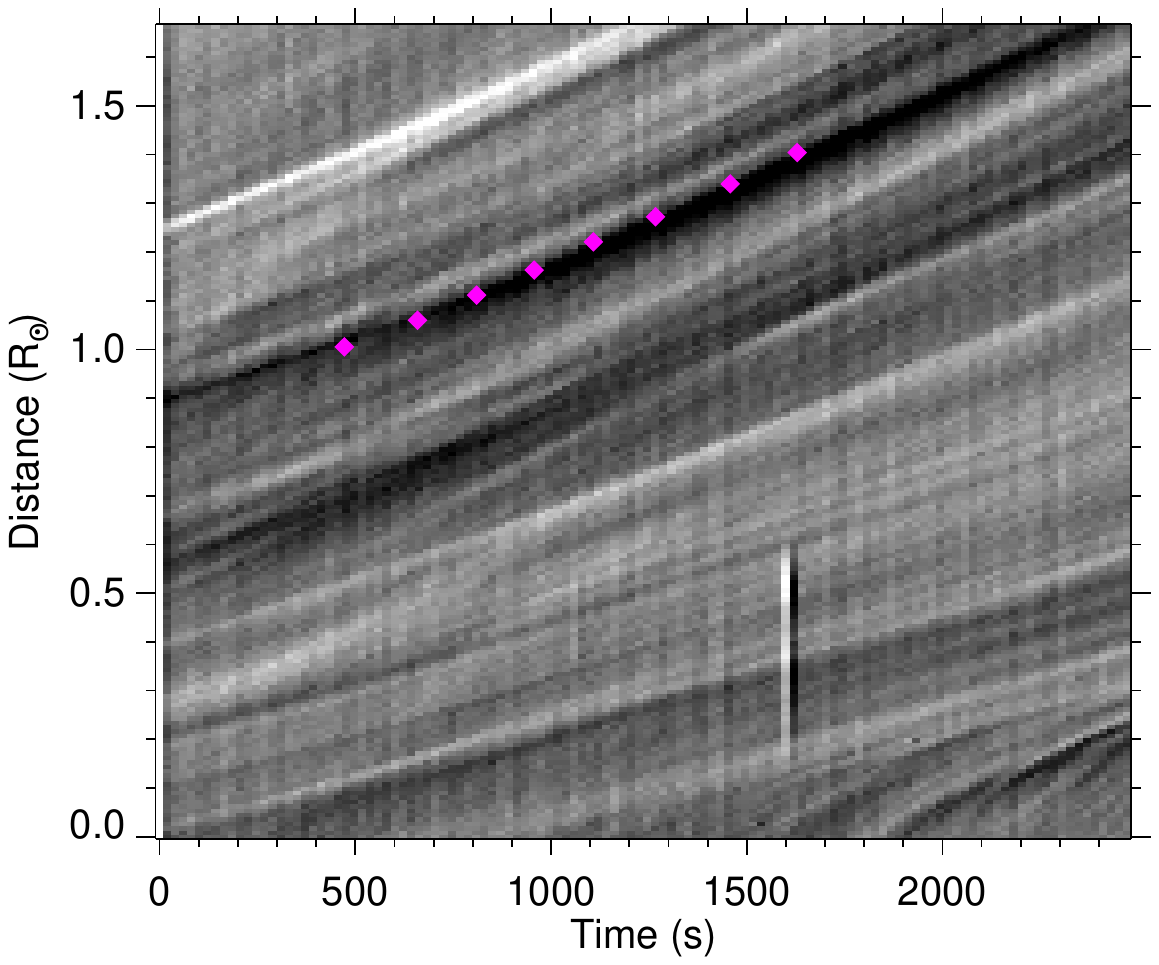}
   \includegraphics[clip, trim=4cm 2cm 3cm 3.cm,width=9.1cm]{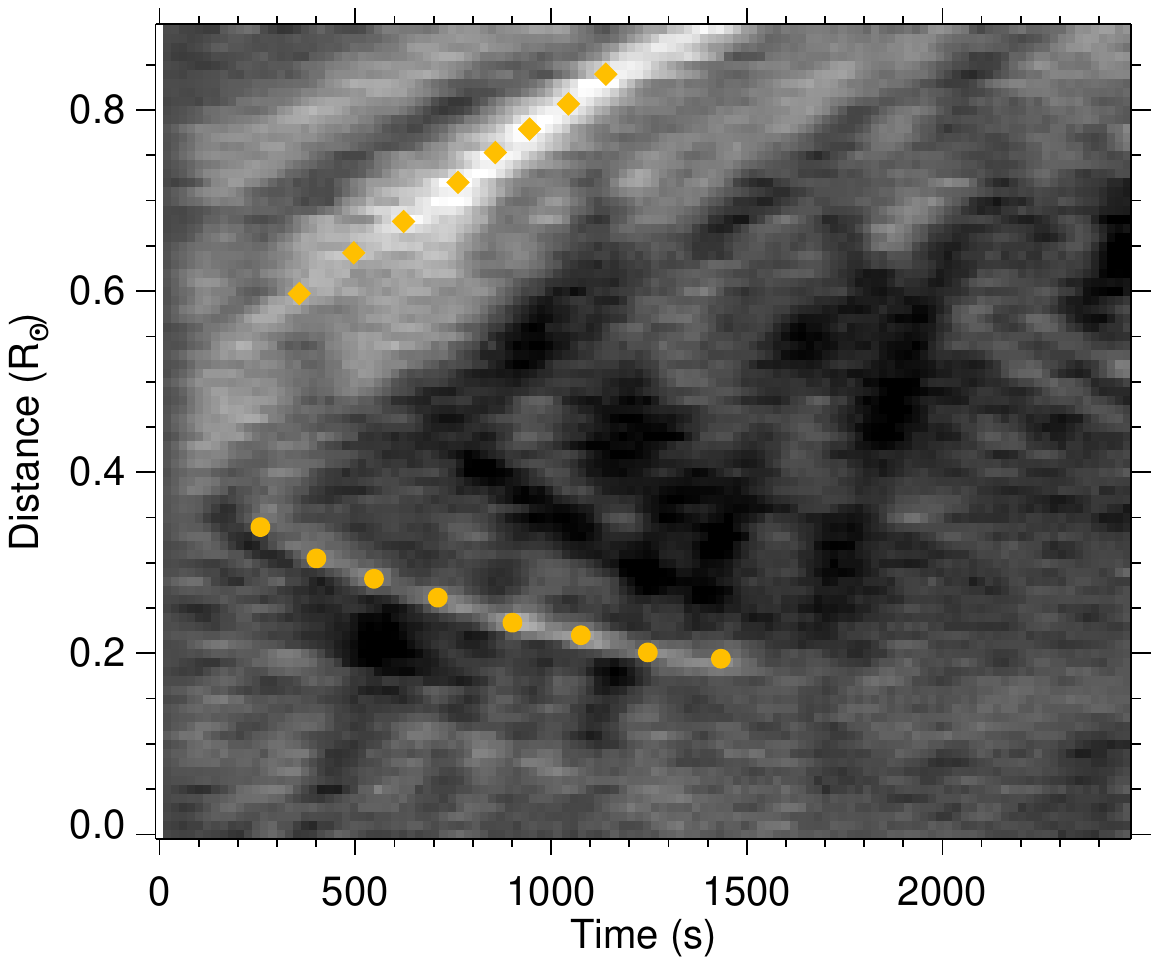}
   \caption{J-maps obtained from the magenta and orange segments in Fig.\,\ref{fig:Metis_HR}. 
   Top panel: J-map extracted along the central position angle of the CME flux rope. It was derived by taking, for each running difference in the sequence, the magenta segment. 
   Bottom panel: J-map obtained crossing the region along the vortex-like descending structure. It was derived from the orange segment by cutting the normalised running difference averaged over a total of five images. Orange diamonds and dots indicate the points used for the polynomial fit.}
              \label{fig:j-maps}%
    \end{figure}

The diagrams illustrated in Fig.\,\ref{fig:j-maps}, corresponding to the magenta A and orange B segments in Fig.\,\ref{fig:Metis_HR}, represent the propagation of the CME's central axis and the trajectory of the vortex-like descending structure, respectively.
In the top panel of Fig.\,\ref{fig:j-maps}, obtained from running difference images, the plasma along the central position angle of the CME flux rope appears as alternating bands of varying grey levels. The speed of the CME front was determined using a second-order fit to the selected magenta diamonds which trace one of the outermost edge of the CME front, yielding a value of speed of approximately 246\,$\pm$\,22\,km/s, at  altitudes between 2 and 3\,R$_\odot$, with an acceleration of 23\,$\pm$\,13\,m/s$^2$.
Using the Carrington coordinates obtained from the 3D reconstruction of the CME front, we deprojected the speed obtaining a radial component of 251\,$\pm$\,22\,km/s. 
The close agreement between these values confirms that the velocity vector at the apex of the CME as estimated from the GCS model (Sec.\,\ref{sec:GCS}) evolved almost entirely within the Metis PoS.  

Despite the availability of only four UV frames, the speed of the dark ring visible in the UV images was derived using the same procedure (the J-map is not shown here, we used a radial cut close to the western flank of the CME). In this case, the deprojected speed was found to be 61\,$\pm$\,12\,km/s, with an acceleration of 34\,$\pm$\,8\,m/s$^2$. This feature was detected about an hour before the high cadence observation sequence of the CME front.

The early phase of the eruption is observable in GOES-R/SUVI 195\,$\r{A}$ images (at $\sim$\,22:30\,UT on October 8$^\mathrm{th}$), where the initial front speed was measured at approximately 16\,$\pm$\,1\,km/s (with slow acceleration of 1.2\,$\pm$\,0.2\,m/s$^2$) at altitudes between 0.2 and 0.5\,R$_\odot$, probably before entering the acceleration phase (see also comments in Sec.\,\ref{subsec:summary})
\footnote{The deprojection was performed assuming a Sun–observer distance of 1 au, consistent with SOHO. While GOES-R is located in geostationary orbit (about 0.01 au farther from the Sun than SOHO), this approximation has a negligible impact on the derived velocities, given the near-disk-center location of the observed features.}.

\begin{figure*}
   \centering
   \includegraphics[clip, trim=6cm 2cm 6cm 3.cm,width=6cm,angle=270]{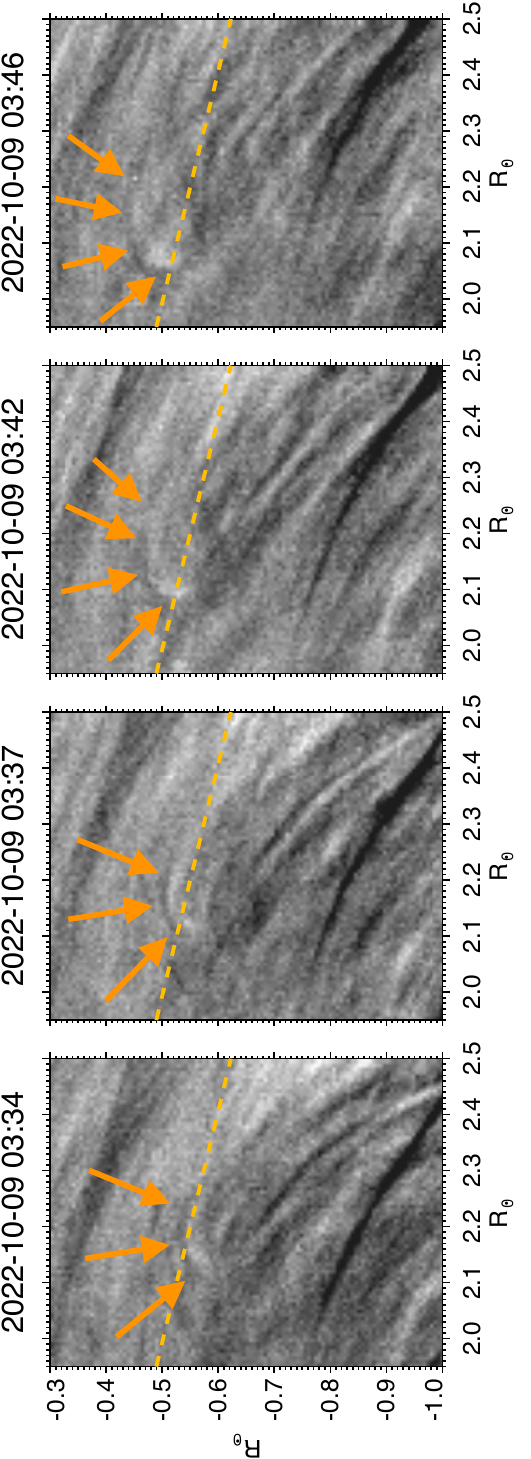}
   \caption{Detailed view of vortex-like structure (feature B) from the high-resolution Metis total brightness observation mode. Orange arrows delineate the rotational dynamics of the plasma, consistent with a vortex morphology. The dashed lines correspond to the orange segment in Fig.\,\ref{fig:Metis_HR}.}
              \label{fig:vortex}%
    \end{figure*}

The bottom panel of Fig.\ref{fig:j-maps} shows the J-map for the vortex-like structure visible in Fig.\ref{fig:Metis_HR} and \ref{fig:vortex} (feature B). In this case, radial segments (orange segment in Fig.\,\ref{fig:Metis_HR} and \ref{fig:vortex}) were extracted from normalised running difference images, averaged over five frames.
The diagram reveals multiple structures with varying slopes in the radial direction, which can be interpreted as plasma upflows and downflows, moving outside the CME circular front.
The speed values on the PoS of these structures were obtained with a second order polinomial fit to the orange points in the bottom panel of Fig.\,\ref{fig:j-maps}. 
The orange dots track the motion of what appears to be the brightest feature of the descending plasma in this region.
Although far from the main CME structure, the upflowing material moves at a PoS speed of 212\,$\pm$\,9\,km/s (from the fit of diamonds in the bottom panel of Fig.\,\ref{fig:j-maps}), comparable to that of the CME front (246\,$\pm$\,22\,km/s). The descending material, on the other hand, moves 30$\%$ more slowly (-138\,$\pm$\,7\,km/s), but exhibits considerable acceleration (102\,$\pm$\,6\,m/s$^2$). 
Furthermore, the evolution observed throughout the image sequence into a circular morphology implies rotational motion or the development of vorticity (see orange arrows in Fig.\,\ref{fig:vortex}).

Since no co-temporal images from other instruments capture these features, deprojection was not possible for features other than the CME front. Therefore, the estimated velocities represent lower limits due to their dependence on the direction of the chosen cut in relation to the actual direction of propagation.

\begin{table}
\centering
      \caption[]{Kinematic parameters of the observed moving features.}
         \label{tab:velocity}
      $ \begin{array}{p{0.25\linewidth}ccc}
            \hline
            \noalign{\smallskip}
            Feature  & Speed & Acceleration    \\
               &   [km/s] &  [m/s^2]    \\
            \noalign{\smallskip}
            \hline
            \noalign{\smallskip}
            Front in Metis (A) & 246\,\pm\,22 &  23\,\pm\,13  \\
              & (251\,\pm\,22) &  (23\,\pm\,13)  \\
            UV dark ring in Metis & 60\,\pm\,12 &  33\,\pm\,8  \\
              & (61\,\pm\,12) &  (34\,\pm\,8)  \\
            Front in SUVI & 15\,\pm\,1 & 1.1\,\pm\,0.2 \\
             & (16\,\pm\,1) & (1.2\,\pm\,0.2) \\
            Downflow (B) & -138\,\pm\,7 &  102\,\pm\,6  \\
            Upflow (B) & 212\,\pm\,9 &  \textnormal{compatible with zero}  \\
            Wave-trains (C) & 482\,\pm\,19 & [-111\div-398]\,\pm\,30  \\
            \noalign{\smallskip}
            \hline
         \end{array} $
         \tablefoot{Speed and acceleration estimates are obtained from the height-time diagrams for the features identified in Metis (high-resolution total brightness and UV) and GOES-R/SUVI 195\,$\r{A}$ imagery.
         The estimates obtained are provided with 1-sigma confidence intervals for the parameters derived from the fit.
         The values in round brackets shown for the CME front and UV dark ring are those obtained by deprojecting the speed on the PoS.}
   \end{table}

\subsection{Coronal wave-trains}
\label{subsec:coronal_waves}

The analysis of high-cadence, normalised running difference images, time-averaged over three (or five) frames, enabled the identification of nearly circular, concentric wavefronts propagating within the Metis FoV, in close proximity to the CME flank. 
These wavefronts originate from the lower part of the CME and propagate along the direction indicated by the green C segment in Fig.\,\ref{fig:Metis_HR} without exhibiting any direct interaction with the erupting plasma. These wavefronts remain visible for almost the entire duration of the 40\,min observational sequence, losing intensity about halfway through the sequence.
Each front appears as successive loop-like waves with bright and dark fronts, with an average width of $\sim$0.08\,R$_\odot$ ($\sim$55\,Mm) per frame. One such wavefront is marked with red diamonds in Fig.\,\ref{fig:Metis_HR}; however, these features are more easily distinguishable in the associated movie Metis\_tB\_HR.mp4 available online, as their brightness remains close to the noise threshold, thus making them difficult to distinguish in individual static frames.

\begin{figure}
   \centering
   \includegraphics[clip, trim=4cm 2cm 3cm 3.cm,width=9.1cm]{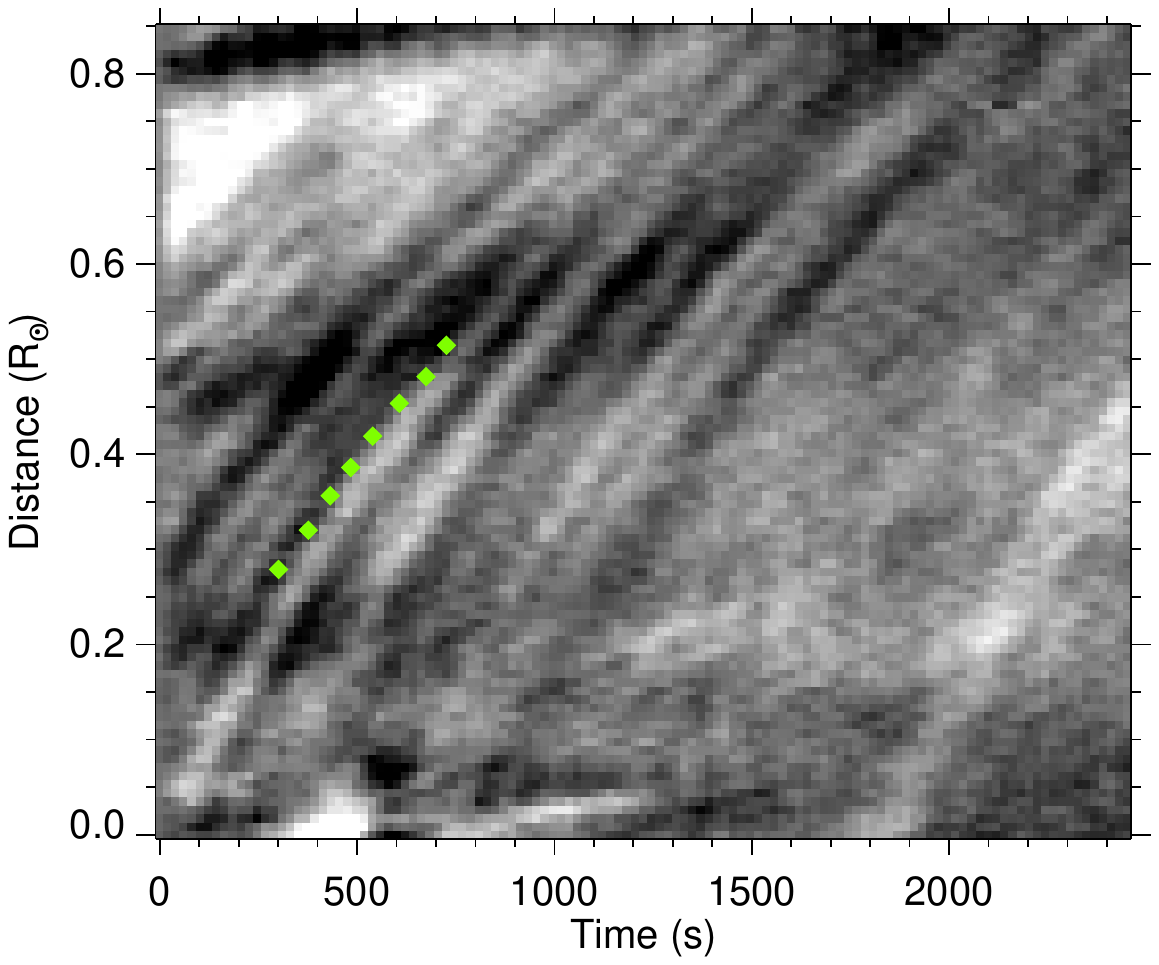}
   \caption{Height-time diagram obtained from the green segment shown in Fig.\,\ref{fig:Metis_HR}, showing the presence of periodic perturbation. The figure was derived by cutting the normalised running difference averaged over a total of three images. Green diamonds indicate one of the points series used for the polynomial fit.}
              \label{fig:j-map_waves}%
    \end{figure}

To improve visibility and infer the wavefront speed, we used the same methodology described in Sec.~\ref{subsec:velocity}. 
In this case, the construction of the height-time diagram require the selection of a segment not in the radial direction but along the propagation direction of the wavefronts. 
A circular fit was performed to one of the wavefronts indicated with red diamonds in Fig.\,\ref{fig:Metis_HR} in order to estimate its centre indicated with the cyan dot. A radial segment passing through this point was selected for analysis. 
The height-time diagram shown in Fig.\,\ref{fig:j-map_waves}, obtained by stacking 40 minutes of green-segment cuts, reveals periodic ridges, which show an inclination following an arc-like pattern.
By applying a second-order polynomial fit to the green diamonds in  Fig.\,\ref{fig:j-map_waves}, we estimate a PoS wavefront group velocity of the order of 482\,$\pm$\,2\,km/s, nearly twice the speed of the CME front (246\,$\pm$\,22\,km/s), with a strong deceleration estimated in the range between -111 and -398\,$\pm$\,30\,m/s$^2$, found from repeated fitting procedures. 

Several factors may contribute to this variability, such as signal-to-noise ratio and structure visibility. 
The wavefronts are close to the noise threshold, making them difficult to track precisely in individual frames. 
The use of time-averaged running difference images enhances the signal, but small variations in background noise fluctuations can introduce uncertainties in the determination of the wavefront position. 
We checked after many attempts that minor shifts in the manual selection of points along the wavefront trajectory could slightly alter the curvature of the second-order fit.

To estimate the frequency of the wavefronts, an empirical analysis based on time series frequency decomposition was employed. These time series were derived by extrapolating the intensity variations over time from height-time diagrams at fixed projected solar radii. 
Two independent techniques were then applied: a wavelet-based approach \citep{wavelets_1998BAMS...79...61T} and the empirical mode decomposition (EMD) algorithm \citep{EMD_1998RSPSA.454..903H}.
At frequencies lower than 10\,mHz, a broad peak was identified around 3\,mHz (corresponding to a $\sim$5\,min periodicity), along with a secondary, more pronounced peak in the 5-6\,mHz range ($\sim$2.7-3\,min periodicity).
The wavelet-based methodology confirmed that both frequency components were statistically significant, with the 5-6\,mHz peak exceeding the 95 per cent confidence level. This frequency range corresponds to the observed frequency of the wavefronts. 

\section{Results and discussion}
\label{sec:results}

\subsection{Summary and analysis of observations}
\label{subsec:summary}

We investigated the CME that occurred on 8–9 October 2022, tracking its evolution from an on-disk filament channel to its propagation through the Metis VL and UV fields of view (see Sec.\,\ref{sec:observation_desc} and Fig.\,\ref{fig:EUI_Metis_tB}). The source region lay along the boundary of a bipolar region, while the formation of a well-defined post-eruptive arcade provides unambiguous evidence of magnetic reconnection beneath the rising flux rope (see Fig.\,\ref{fig:SUVI} and \ref{fig:GONG}). The absence of a detectable flare demonstrates that the reconnection was weak and released limited energy.

In the low corona (0.2-0.5\,R$_\odot$), the CME front expanded slowly, with radial speed below 20\,km/s with minimal acceleration, derived from expanding loops.  
At $\sim$1.8\,R$_\odot$ from the solar photosphere, a well-defined dark ring structure was detected in the Metis UV channel (see Fig.\,\ref{fig:EUI_Metis_tB}).
This structure, faintly visible in GOES-R/SUVI 195\,$\r{A}$ (due to the different point of view) and SolO/EUI-FSI 174\,$\r{A}$ images, corresponds to the CME front propagating through regions characterized by lower temperatures and higher densities, in agreement with previous characterisations of slow CME fronts \citep{Robbrecht_stealth_2009ApJ...701..283R,D'Huys_stealth_2014ApJ...795...49D}. 

At greater heights, between 2 and 3\,R$_\odot$, multi-viewpoint 3D reconstruction yields a CME front deprojected speed of 251\,$\pm$\,22\,km/s and an acceleration of 23\,$\pm$\,13\,m/s$^2$, with propagation direction strongly aligned with the Metis plane of sky (see Sec.\,\ref{sec:GCS}). 
These values confirm the identification of the event as a slow CME, still in its acceleration phase within the 2\,R$_\odot$ range near the Metis FoV outer limit. This finding is consistent with previous studies on the kinematic evolution of slow-CMEs \citep{Temmer_kinem_CME_2016AN....337.1010T} as well as with statistical observations of CMEs during solar minimum using SOHO/LASCO-C2  \citep{Yashiro_CME_SOHO_catalogue_2004JGRA..109.7105Y}.

The observations suggest that the event shares some characteristics that are commonly associated with stealth CMEs \citep{Robbrecht_stealth_2009ApJ...701..283R,Howard_stealth_2013SoPh..285..269H,Lynch_stealth_2016JGRA..12110677L}.
Classical low-coronal signatures were weak or absent, and the eruption originated from a filament channel, a region of sheared magnetic field aligned with the PIL and with modest magnetic fields.
Similar events were discussed in works such as \citealp{Palmerio_2021FrASS...8..109P} and \citealp{O'Kane_2021ApJ...908...89O}.
 
Finally, Metis high-cadence total brightness observations also revealed the presence of two notable features: a structure resembling a vortex descending along the flank of the CME and circular coronal wave-trains (so far mostly observed in EUV) propagating ahead of the CME front  (see Sec.\,\ref{subsec:velocity} and \ref{subsec:coronal_waves}). Their detection underscores the diagnostic power of Metis for resolving fine-scale CME dynamics in the middle corona and will be discussed in detail in the following sections.

\subsection{On the vortex-like feature}
\label{subsec:KHi}

As illustrated in Fig.\,\ref{fig:Metis_HR} and in Fig.\,\ref{fig:vortex}, a large descending structure resembling a vortex is observed moving along the western flank of the CME. 
This motion may result from frictional interaction between the expanding CME plasma and the ambient corona. 

Previous observations from SOHO/LASCO and STEREO/SECCHI coronagraphs \citep{Sheeley_Wang_inflows_2002ApJ...579..874S,Tripathi_magn_reconn_2006A&A...453.1111T,Tripathi_downflow_2007A&A...472..633T} have documented return flows of plasma trailing CMEs, driven by a combination of gravitational forces and magnetic tension. 
In our case, the descending material may represent such a return flow, initiated after the plasma loses outward momentum, possibly due to interaction with overlying coronal structures or field-line retractions \citep{Tripathi_downflow_2007A&A...472..633T}. 
The fact that the upflowing plasma maintains a velocity comparable to that of the CME suggests that it remains coupled to the expanding eruption, while the descending component becomes decoupled and decelerated by external forces.

A velocity-shear-driven Kelvin-Helmholtz instability (KHi) presents a plausible alternative mechanism for the structural evolution observed here.
These occur at the interface of two plasma flows with different velocities, and are known to produce vortex-like structures due to shear-driven turbulence.
In the context of solar physics, the KHi have been observed in multiple studies along the edges of CMEs (i.e. \citealp{Foullon_KH_2011ApJ...729L...8F,Mostl_KHi_2013ApJ...766L..12M,Paouris_KH_PSP_2024ApJ...964..139P,Ofman_KHi_CME_2026ApJ...997L..28O}).
In solar observations, KH instabilities have been reported along CME boundaries as trains of small vortices, typically with projected regular separation distances of tens of Mm at heights below 150\,Mm. 
The propagation speed of the vortices envelope on the PoS (referred to as group speed), is approximately half the speed of the ejecta front, consistent with predictions from linear theory \citep{Chandrasekhar}.
The structure observed in this event, however, differs from classical vortex trains: it exhibits a single, large, descending vortex with a PoS speed of -138\,$\pm$\,7\,km/s (group speed on the PoS with a speed 30\% lower than the CME front as discussed in Sec.\,\ref{subsec:velocity}) and a strong acceleration (see Tab.\,\ref{tab:velocity}), spanning a projected  linear size of $\sim$80\,Mm ($\sim$0.11\,R$_\odot$) in a nearly concentric loop shape. 
These features are consistent with turbulence at the CME flank, potentially driven by the KHi. 
If confirmed, this would represent one of the first detections of a large-scale KHi at middle-coronal heights. 
A similar KHi-like feature was hypothesised for the CME observed on 2 October 2021 by Metis, particularly bright in the UV channel in the work of \cite{Russano_CME_UV_2024A&A...683A.191R}.

The orientation and strength of the magnetic field plays a key role in the evolution of KHi. As demonstrated in theoretical studies \citep{Chandrasekhar}, strong magnetic fields can suppress KHi growth, while weaker or sheared fields may allow larger and more complex vortex formation. Although the magnetic field configuration in this event cannot be directly assessed, it is plausible that sheared structures along the CME boundary contributed to the observed vortex morphology.
However, detailed characterization of the magnetic field in this event is beyond the scope of the present work. A future investigation could benefit from magnetohydrodynamic (MHD) simulation similar to those performed by \cite{Syntelis_KH_sim_2019ApJ...884L...4S} to assess the stability of the CME flank under velocity shear conditions.

\subsection{Quasi-periodic wave trains discussion}
\label{subsec:QPF}

The coronal wave-trains analysed in Sec.\,\ref{subsec:coronal_waves} can be interpreted in the light of recent studies on quasi-periodic wave trains or quasi-periodic fast-modes (QFPs) (i.e. \citealp{Liu_QPF_2012ApJ...753...52L} and \citealp{Shen_QPF_2022SoPh..297...20S}).
These wave trains are typically observed in the low corona by SDO/AIA, and are characterized by coherent, concentric wavefronts propagating outward, often associated with flares and EUV waves.
They travel along or across coronal loops at speeds ranging from several hundred to over 2000 km/s, and typically display periodicities of a few minutes with measurable deceleration.

As summarized in Tab.\,\ref{tab:waves}, the coronal wave-trains observed by Metis share some similarities with the QFPs wave trains described by \cite{Liu_QPF_2012ApJ...753...52L} and \cite{Shen_QPF_2022SoPh..297...20S} (such as their coherence, circular shape, and periodicity) but also exhibit notable differences.
In particular, the Metis wavefronts propagate at lower speeds ($\sim$500\,km/s) compared to the $\sim$1400\,km/s for AIA QFPs, show strong deceleration possibly due to the
higher coronal heights they are observed (2-3\,R$_\odot$). They are not associated with any detectable flare or EUV wave signature.
These differences likely reflect the distinct coronal environment in which the Metis wavefronts evolve.

The observed frequency range (5-6\,mHz or period of 2.7–3\,minutes) is consistent with QFPs, and falls within the magnetoacoustic regime.
Other oscillatory modes observed in this frequency range include  the well-known \(p\)-modes, which has been shown to propagate in the upper solar atmosphere, wherein magnetic forces dominate \citep{Jess_wave_review_2023LRSP...20....1J}.
However, the coronal wave-trains detected by Metis are distinguished by their spatial confinement within an angular extent of roughly 50$^\circ$, alongside a limited temporal duration, as evidenced by a progressive decrease in brightness throughout the sequence. This behavior is indicative of an impulsive triggered phenomenon. 

\begin{table}
\centering
      \caption[]{Summary of observational and kinematic parameters for SDO/AIA and Metis wave trains.}
         \label{tab:waves}
        \begin{tabular}{l c c}
            \hline
            \noalign{\smallskip}
            Observations  &  SDO/AIA  & SolO/Metis     \\
             & wave trains & coronal wave-trains \\ 
            \noalign{\smallskip}
            \hline
            \noalign{\smallskip}
            Heliocentric height & < 1.5 R$_\odot$ &  Whitin 3 R$_\odot$  \\
            Speed & $\sim$ 1400 km/s  &  $\sim$ 500 km/s   \\
             &  decelerating    & decelerating \\
              & up to 650 km/s & \\
            Period  & 2 min &  2.7 - 3 min  \\
            CME speed & 250 km/s &  251 $\pm$ 3 km/s  \\
            Lateral expansion & Yes &  Yes \\
            EUV wave & Yes & No \\
            Flare & Yes & No  \\
            Coherent  & R$_\odot$/2 & R$_\odot$/2 \\
            travel distance & & \\
            \noalign{\smallskip}
            \hline
         \end{tabular}
         \tablefoot{Observational parameters for SDO/AIA wave trains are adapted from \citep{Liu_QPF_2012ApJ...753...52L}. Characteristics for Metis coronal wave-trains are detailed in Section,\ref{subsec:coronal_waves}.}
   \end{table}

A key difference observed in this study is the lack of an associated EUV wave or flare, which is typically observed in 80$\%$ of QFP wave train events \citep{Shen_QPF_2022SoPh..297...20S}.
Most observational and theoretical studies as well as models and MHD simulations pertaining to QFP modes have attributed QFP wave trains to flare-associated periodic drivers, where intensity oscillations are observed across a wide wavelength range, from radio to $\gamma$-rays. 
Instead, we propose that the Metis wavefronts are generated by CME-related mechanisms, such as lateral expansion of the CME envelope, producing thin, pulse-like perturbations \citep{Patsourakos_QFP_Ex_2010A&A...522A.100P}, or sequential stretching of reconnected magnetic field lines, as the flux rope erupts and expands \citep{Shen_QPF_2022SoPh..297...20S,Shen_QFP_ex_2022A&A...665A..51S}.

The QFP waves identified by \cite{Liu_QPF_2012ApJ...753...52L} have been recently replicated in numerical simulations by \cite{Hu_sim_QFP_2024ApJ...962...42H}. These authors propose that large-scale QFP waves, observed on the flank of the CME, are excited by internal disturbances within the flux rope triggered by a loss of equilibrium. According to this model, a fraction of the internal perturbation leaks through the flux rope boundary into the ambient corona, manifesting as the observed QFP wave trains.
This interpretation could also explain the circular wavefronts observed by Metis, which seem to originate from within the CME envelope in the SolO/EUI-FSI field of view (see red diamonds indicating one of the circular fronts with its centre denoted by the cyan dot at a radial distance of 1.6\,R$_\odot$ in Fig.\,\ref{fig:Metis_HR}). 
As outlined in Sec.\,\ref{sec:morphology} and in accordance with the model proposed by \cite{Sun_magn_reconn_2015NatCo...6.7598S}, this region may represent the magnetic reconnection zone responsible for the eruption of the flux rope and, on the lateral side, of this impulsive phenomenon.
The SolO/EUI-FSI, GOES/SUVI and SDO/AIA data were also checked for wave-like structures by applying the algorithm presented in Sec.\,\ref{sec:HR_res}. In the first two cases the cadence is too high, while in the third case the cadence is 12\,s (compatible with Metis VL), however the detector background noise in the lower corona is too high to obtain good detection and the waves probably form higher up in the corona.

We cannot exclude the possibility that a weak or undetected flare occurred, as observational limitations may have obscured signatures \citep{Chen_CME_review_2011LRSP....8....1C}. Nonetheless, this event provides a compelling example of CME-driven quasi-periodic wave trains in the middle corona, independent of flare-associated drivers. 
As such, it offers a rare opportunity to study wave generation mechanisms in the extended corona and possibly observe for the first time in the high-corona the site where the process of disconnection happens, under conditions not typically accessible with low-coronal imagers.
This might give insight to evolution of stealth CMEs.

Additionally, a search for radio signatures was conducted within the relevant time range to identify potential shock waves or electron beams at low coronal heights (below 2\,R$_\odot$), but none were found 
due to the lack of simultaneous radio heliograph observations.

\section{Conclusions}
\label{sec:conclusions}

This work has demonstrated the unique capability of the Metis coronagraph onboard Solar Orbiter to capture high-cadence and high-resolution observations, enabling unprecedented insights into transient coronal phenomena.
The detailed analysis of the 8–9 October 2022 event revealed several important findings:
\begin{itemize}
\item Successful capture of the intricate CME morphology, unveiling fine-scale internal structures (from the highly structured, non-uniform leading edge to the dense core) not achievable with lower temporal resolution of other space-based coronagraphs.
\item  Identification of a prominent, descending vortex-like structure observed along the CME flank, that could be interpreted as evidence of a Kelvin–Helmholtz instability. If confirmed, it would constitute one of the first reported detections of large-scale KHi at intermediate coronal altitudes.
\item High-resolution detection of quasi-periodic coronal wavefronts, attributed to fast-mode magnetosonic waves or CME-driven field-line oscillations. Their occurrence without flare or EUV wave signatures suggests a dominant role of CME expansion and reconnection, with Metis wavefronts differing from classical QFPs in speed, altitude, and excitation mechanisms.
\end{itemize}

The event studied here serves as a valuable test case for modeling wave excitation and instability onset in the extended corona, and underscores the importance of coordinated, multi-instrument, multi-height analyses for fully understanding CME propagation and coronal response.

Future coordinated campaigns combining Metis with instruments such as SolO/EUI, SolO/STIX, Parker Solar Probe, SDO and ground-based radio observatories will be crucial for further constraining the physical mechanisms behind these phenomena.
Additionally, MHD simulations will be essential to validate the interpretation of wave generation, vortex evolution and magnetic field reconfiguration in CME events observed at these heights.

This study also opens new avenues for discussion, including the role of stealth CMEs (slow, low-energy eruptions lacking classical low-coronal signatures) in driving wave and instability phenomena in the middle corona. Understanding these subtle eruptions is crucial for improving space weather forecasting, given their potential geoeffectiveness despite the absence of precursors.

In summary, Metis observations provide a transformative perspective on CME-driven dynamics in the middle corona, offering new insights into wave phenomena, plasma instabilities, and magnetic restructuring processes that shape the solar corona’s response to eruptive events.

\begin{acknowledgements}
    Solar Orbiter is a space mission of international collaboration between ESA and NASA, operated by ESA.  Metis was built and operated with funding from the Italian Space Agency (ASI), under contracts to the National Institute of Astrophysics (INAF) and industrial partners. Metis was built with hardware contributions from Germany (Bundesministerium für Wirtschaft und Energie through DLR), from the Czech Republic (PRODEX) and from ESA.  
    Metis team thanks the former PI, Ester Antonucci, for leading the development of Metis until the final delivery to ESA. \href{ https://doi.org/10.5270/esa-366ut35}{Metis landing page.}
    F. Frassati acknowledges support from INAF grant CUP C63C23000810005 "IDEA-SW - Integrating Data and Expertise to Advance Space Weather forecasting of Catastrophic Events" and from the Project “Supporto per la realizzazione degli strumenti Metis, SWA DPU e STIX” CUP F86C18000570005.  
    Y.D.L. was partially supported by  the Austrian Science Fund (FWF) 10.55776/ESP2660425.
    Data were acquired by GONG instruments operated by NISP/NSO/AURA/NSF with contribution from NOAA.
    The authors thank Dr. Z. Liang from the Max Planck Institute for Solar System Research in G\"oettingen for the valuable support that contributed to this work.
    Wavelet software was provided by C. Torrence and G. Compo, and is available at URL: http://paos.colorado.edu/research/wavelets/
    The EUI instrument was built by CSL, IAS, MPS, MSSL/UCL, PMOD/WRC, ROB, LCF/IO with funding from the Belgian Federal Science Policy Office (BELSPO/PRODEX PEA 4000112292 and 4000134088); the Centre National d’Etudes Spatiales (CNES); the UK Space Agency (UKSA); the Bundesministerium für Wirtschaft und Energie (BMWi) through the Deutsches Zentrum für Luft- und Raumfahrt (DLR); and the Swiss Space Office (SSO). 
    The GOES 16 X-ray / magnetic field / particle data are produced in real time by the NOAA Space Weather Prediction Center (SWPC) and are distributed by the NOAA National Geophysical Data Center (NGDC).
    This work utilizes data produced collaboratively between AFRL/ADAPT and NSO/NISP.
\end{acknowledgements}

\bibliographystyle{aa} 
\bibliography{biblio_main}

@ARTICLE{metis_instr,
       author = {{Antonucci}, Ester and {Romoli}, Marco and {Andretta}, Vincenzo and {Fineschi}, Silvano and {Heinzel}, Petr and {Moses}, J. Daniel and {Naletto}, Giampiero and {Nicolini}, Gianalfredo and {Spadaro}, Daniele and {Teriaca}, Luca and {Berlicki}, Arkadiusz and {Capobianco}, Gerardo and {Crescenzio}, Giuseppe and {Da Deppo}, Vania and {Focardi}, Mauro and {Frassetto}, Fabio and {Heerlein}, Klaus and {Landini}, Federico and {Magli}, Enrico and {Marco Malvezzi}, Andrea and {Massone}, Giuseppe and {Melich}, Radek and {Nicolosi}, Piergiorgio and {Noci}, Giancarlo and {Pancrazzi}, Maurizio and {Pelizzo}, Maria G. and {Poletto}, Luca and {Sasso}, Clementina and {Sch{\"u}hle}, Udo and {Solanki}, Sami K. and {Strachan}, Leonard and {Susino}, Roberto and {Tondello}, Giuseppe and {Uslenghi}, Michela and {Woch}, Joachim and {Abbo}, Lucia and {Bemporad}, Alessandro and {Casti}, Marta and {Dolei}, Sergio and {Grimani}, Catia and {Messerotti}, Mauro and {Ricci}, Marco and {Straus}, Thomas and {Telloni}, Daniele and {Zuppella}, Paola and {Auch{\`e}re}, Frederic and {Bruno}, Roberto and {Ciaravella}, Angela and {Corso}, Alain J. and {Alvarez Copano}, Miguel and {Aznar Cuadrado}, Regina and {D'Amicis}, Raffaella and {Enge}, Reiner and {Gravina}, Alessio and {Jej{\v{c}}i{\v{c}}}, Sonja and {Lamy}, Philippe and {Lanzafame}, Alessandro and {Meierdierks}, Thimo and {Papagiannaki}, Ioanna and {Peter}, Hardi and {Fernandez Rico}, German and {Giday Sertsu}, Mewael and {Staub}, Jan and {Tsinganos}, Kanaris and {Velli}, Marco and {Ventura}, Rita and {Verroi}, Enrico and {Vial}, Jean-Claude and {Vives}, Sebastien and {Volpicelli}, Antonio and {Werner}, Stephan and {Zerr}, Andreas and {Negri}, Barbara and {Castronuovo}, Marco and {Gabrielli}, Alessandro and {Bertacin}, Roberto and {Carpentiero}, Rita and {Natalucci}, Silvia and {Marliani}, Filippo and {Cesa}, Marco and {Laget}, Philippe and {Morea}, Danilo and {Pieraccini}, Stefano and {Radaelli}, Paolo and {Sandri}, Paolo and {Sarra}, Paolo and {Cesare}, Stefano and {Del Forno}, Felice and {Massa}, Ernesto and {Montabone}, Mauro and {Mottini}, Sergio and {Quattropani}, Daniele and {Schillaci}, Tiziano and {Boccardo}, Roberto and {Brando}, Rosario and {Pandi}, Arianna and {Baietto}, Cristian and {Bertone}, Riccardo and {Alvarez-Herrero}, Alberto and {Garc{\'\i}a Parejo}, Pilar and {Cebollero}, Mar{\'\i}a and {Amoruso}, Mauro and {Centonze}, Vito},
        title = "{Metis: the Solar Orbiter visible light and ultraviolet coronal imager}",
      journal = {\aap},
         year = 2020,
        month = oct,
       volume = {642},
          eid = {A10},
        pages = {A10},
          doi = {10.1051/0004-6361/201935338},
archivePrefix = {arXiv},
       eprint = {1911.08462},
 primaryClass = {astro-ph.SR},
       adsurl = {https://ui.adsabs.harvard.edu/abs/2020A&A...642A..10A}
}

@ARTICLE{LASCO_1995SoPh..162..357B,
       author = {{Brueckner}, G.~E. and {Howard}, R.~A. and {Koomen}, M.~J. and {Korendyke}, C.~M. and {Michels}, D.~J. and {Moses}, J.~D. and {Socker}, D.~G. and {Dere}, K.~P. and {Lamy}, P.~L. and {Llebaria}, A. and {Bout}, M.~V. and {Schwenn}, R. and {Simnett}, G.~M. and {Bedford}, D.~K. and {Eyles}, C.~J.},
        title = "{The Large Angle Spectroscopic Coronagraph (LASCO)}",
      journal = {\solphys},
         year = 1995,
        month = dec,
       volume = {162},
       number = {1-2},
        pages = {357-402},
          doi = {10.1007/BF00733434},
       adsurl = {https://ui.adsabs.harvard.edu/abs/1995SoPh..162..357B}
}

@ARTICLE{SOHO_1995SoPh..162....1D,
       author = {{Domingo}, V. and {Fleck}, B. and {Poland}, A.~I.},
        title = "{The SOHO Mission: an Overview}",
      journal = {\solphys},
         year = 1995,
        month = dec,
       volume = {162},
       number = {1-2},
        pages = {1-37},
          doi = {10.1007/BF00733425},
       adsurl = {https://ui.adsabs.harvard.edu/abs/1995SoPh..162....1D}
}

@ARTICLE{STEREO_2008SSRv..136....5K,
       author = {{Kaiser}, M.~L. and {Kucera}, T.~A. and {Davila}, J.~M. and {St. Cyr}, O.~C. and {Guhathakurta}, M. and {Christian}, E.},
        title = "{The STEREO Mission: An Introduction}",
      journal = {\ssr},
         year = 2008,
        month = apr,
       volume = {136},
       number = {1-4},
        pages = {5-16},
          doi = {10.1007/s11214-007-9277-0},
       adsurl = {https://ui.adsabs.harvard.edu/abs/2008SSRv..136....5K}
}

@ARTICLE{SolO_2020A&A...642A...1M,
       author = {{M{\"u}ller}, D. and {St. Cyr}, O.~C. and {Zouganelis}, I. and {Gilbert}, H.~R. and {Marsden}, R. and {Nieves-Chinchilla}, T. and {Antonucci}, E. and {Auch{\`e}re}, F. and {Berghmans}, D. and {Horbury}, T.~S. and {Howard}, R.~A. and {Krucker}, S. and {Maksimovic}, M. and {Owen}, C.~J. and {Rochus}, P. and {Rodriguez-Pacheco}, J. and {Romoli}, M. and {Solanki}, S.~K. and {Bruno}, R. and {Carlsson}, M. and {Fludra}, A. and {Harra}, L. and {Hassler}, D.~M. and {Livi}, S. and {Louarn}, P. and {Peter}, H. and {Sch{\"u}hle}, U. and {Teriaca}, L. and {del Toro Iniesta}, J.~C. and {Wimmer-Schweingruber}, R.~F. and {Marsch}, E. and {Velli}, M. and {De Groof}, A. and {Walsh}, A. and {Williams}, D.},
        title = "{The Solar Orbiter mission. Science overview}",
      journal = {\aap},
         year = 2020,
        month = oct,
       volume = {642},
          eid = {A1},
        pages = {A1},
          doi = {10.1051/0004-6361/202038467},
archivePrefix = {arXiv},
       eprint = {2009.00861},
 primaryClass = {astro-ph.SR},
       adsurl = {https://ui.adsabs.harvard.edu/abs/2020A&A...642A...1M}
}

@ARTICLE{DeLeo_VL_2023A&A...676A..45D,
       author = {{De Leo}, Y. and {Burtovoi}, A. and {Teriaca}, L. and {Romoli}, M. and {Chioetto}, P. and {Andretta}, V. and {Uslenghi}, M. and {Landini}, F. and {Susino}, R. and {Pancrazzi}, M. and {Frassati}, F. and {Giarrusso}, M. and {Giordano}, S. and {Zangrilli}, L. and {Spadaro}, D. and {Abbo}, L. and {Bemporad}, A. and {Capobianco}, G. and {Capuano}, G.~E. and {Casini}, C. and {Casti}, M. and {Corso}, A.~J. and {Da Deppo}, V. and {Fabi}, M. and {Fineschi}, S. and {Frassetto}, F. and {Grimani}, C. and {Guglielmino}, S.~L. and {Heinzel}, P. and {Jerse}, G. and {Liberatore}, A. and {Magli}, E. and {Massone}, G. and {Messerotti}, M. and {Moses}, J.~D. and {Naletto}, G. and {Nicolini}, G. and {Pelizzo}, M.~G. and {Romano}, P. and {Russano}, G. and {Sasso}, C. and {Sch{\"u}hle}, U. and {Straus}, T. and {Slemer}, A. and {Stangalini}, M. and {Telloni}, D. and {Volpicelli}, C.~A. and {Zuppella}, P.},
        title = "{In-flight radiometric calibration of the Metis Visible Light channel using stars and comparison with STEREO-A/COR2 data}",
      journal = {\aap},
         year = 2023,
        month = aug,
       volume = {676},
          eid = {A45},
        pages = {A45},
          doi = {10.1051/0004-6361/202345979},
       adsurl = {https://ui.adsabs.harvard.edu/abs/2023A&A...676A..45D}
}

@ARTICLE{DeLeo_UV_2025A&A...697A..73D,
       author = {{De Leo}, Y. and {Burtovoi}, A. and {Teriaca}, L. and {Romoli}, M. and {Andretta}, V. and {Uslenghi}, M. and {Giordano}, S. and {Chioetto}, P. and {Susino}, R. and {Landini}, F. and {Pancrazzi}, M. and {Frassati}, F. and {Russano}, G. and {Spadaro}, D. and {Abbo}, L. and {Bemporad}, A. and {Capobianco}, G. and {Capuano}, G. and {Casini}, C. and {Casti}, M. and {Corso}, A.~J. and {Da Deppo}, V. and {Fabi}, M. and {Fineschi}, S. and {Frassetto}, F. and {Giarrusso}, M. and {Grimani}, C. and {Guglielmino}, S.~L. and {Heinzel}, P. and {Jerse}, G. and {Liberatore}, A. and {Magli}, E. and {Massone}, G. and {Messerotti}, M. and {Moses}, J.~D. and {Naletto}, G. and {Nicolini}, G. and {Pelizzo}, M.~G. and {Romano}, P. and {Sasso}, C. and {Sch{\"u}hle}, U. and {Straus}, T. and {Slemer}, A. and {Stangalini}, M. and {Telloni}, D. and {Volpicelli}, C.~A. and {Zangrilli}, L. and {Zuppella}, P.},
        title = "{In-flight radiometric calibration of the Metis UV H I Ly-{\ensuremath{\alpha}} channel and comparison with UVCS data}",
      journal = {\aap},
         year = 2025,
        month = may,
       volume = {697},
          eid = {A73},
        pages = {A73},
          doi = {10.1051/0004-6361/202450759},
       adsurl = {https://ui.adsabs.harvard.edu/abs/2025A&A...697A..73D}
}

@ARTICLE{MGN,
       author = {{Morgan}, Huw and {Druckm{\"u}ller}, Miloslav},
        title = "{Multi-Scale Gaussian Normalization for Solar Image Processing}",
      journal = {\solphys},
         year = 2014,
        month = aug,
       volume = {289},
       number = {8},
        pages = {2945-2955},
          doi = {10.1007/s11207-014-0523-9},
archivePrefix = {arXiv},
       eprint = {1403.6613},
 primaryClass = {astro-ph.SR},
       adsurl = {https://ui.adsabs.harvard.edu/abs/2014SoPh..289.2945M}
}

@ARTICLE{Chen_CME_review_2011LRSP....8....1C,
       author = {{Chen}, P.~F.},
        title = "{Coronal Mass Ejections: Models and Their Observational Basis}",
      journal = {Living Reviews in Solar Physics},
         year = 2011,
        month = apr,
       volume = {8},
       number = {1},
          eid = {1},
        pages = {1},
          doi = {10.12942/lrsp-2011-1},
       adsurl = {https://ui.adsabs.harvard.edu/abs/2011LRSP....8....1C}
}

@ARTICLE{Webb_CME_review_2012LRSP....9....3W,
       author = {{Webb}, David F. and {Howard}, Timothy A.},
        title = "{Coronal Mass Ejections: Observations}",
      journal = {Living Reviews in Solar Physics},
         year = 2012,
        month = jun,
       volume = {9},
       number = {1},
          eid = {3},
        pages = {3},
          doi = {10.12942/lrsp-2012-3},
       adsurl = {https://ui.adsabs.harvard.edu/abs/2012LRSP....9....3W}
}

@ARTICLE{post_eruption_loop_Tripathi2004A&A...422..337T,
       author = {{Tripathi}, D. and {Bothmer}, V. and {Cremades}, H.},
        title = "{The basic characteristics of EUV post-eruptive arcades and their role as tracers of coronal mass ejection source regions}",
      journal = {\aap},
         year = 2004,
        month = jul,
       volume = {422},
        pages = {337-349},
          doi = {10.1051/0004-6361:20035815},
       adsurl = {https://ui.adsabs.harvard.edu/abs/2004A&A...422..337T}
}

@ARTICLE{cremades_3D_CME_2004A&A...422..307C,
       author = {{Cremades}, H. and {Bothmer}, V.},
        title = "{On the three-dimensional configuration of coronal mass ejections}",
      journal = {\aap},
         year = 2004,
        month = jul,
       volume = {422},
        pages = {307-322},
          doi = {10.1051/0004-6361:20035776},
       adsurl = {https://ui.adsabs.harvard.edu/abs/2004A&A...422..307C}
}

@ARTICLE{Howard_CME_shape_on_POS_2017ApJ...834...86H,
       author = {{Howard}, T.~A. and {DeForest}, C.~E. and {Schneck}, U.~G. and {Alden}, C.~R.},
        title = "{Challenging Some Contemporary Views of Coronal Mass Ejections. II. The Case for Absent Filaments}",
      journal = {\apj},
         year = 2017,
        month = jan,
       volume = {834},
       number = {1},
          eid = {86},
        pages = {86},
          doi = {10.3847/1538-4357/834/1/86},
       adsurl = {https://ui.adsabs.harvard.edu/abs/2017ApJ...834...86H}
}

@ARTICLE{Russano_CME_UV_2024A&A...683A.191R,
       author = {{Russano}, G. and {Andretta}, V. and {De Leo}, Y. and {Teriaca}, L. and {Uslenghi}, M. and {Giordano}, S. and {Telloni}, D. and {Heinzel}, P. and {Jej{\v{c}}i{\v{c}}}, S. and {Abbo}, L. and {Bemporad}, A. and {Burtovoi}, A. and {Capuano}, G.~E. and {Frassati}, F. and {Guglielmino}, S.~L. and {Jerse}, G. and {Landini}, F. and {Liberatore}, A. and {Nicolini}, G. and {Pancrazzi}, M. and {Romano}, P. and {Sasso}, C. and {Susino}, R. and {Zangrilli}, L. and {Da Deppo}, V. and {Fineschi}, S. and {Grimani}, C. and {Moses}, J.~D. and {Naletto}, G. and {Romoli}, M. and {Spadaro}, D. and {Stangalini}, M.},
        title = "{Eruptive events with exceptionally bright emission in H I Ly-{\ensuremath{\alpha}} observed by the Metis coronagraph}",
      journal = {\aap},
         year = 2024,
        month = mar,
       volume = {683},
          eid = {A191},
        pages = {A191},
          doi = {10.1051/0004-6361/202347741},
archivePrefix = {arXiv},
       eprint = {2312.01899},
 primaryClass = {astro-ph.SR},
       adsurl = {https://ui.adsabs.harvard.edu/abs/2024A&A...683A.191R}
}

@ARTICLE{Howard_CME_history_2023FrASS..1064226H,
       author = {{Howard}, Russell A. and {Vourlidas}, Angelos and {Stenborg}, Guillermo},
        title = "{The evolution of our understanding of coronal mass ejections}",
      journal = {Frontiers in Astronomy and Space Sciences},
         year = 2023,
        month = nov,
       volume = {10},
          eid = {1264226},
        pages = {1264226},
          doi = {10.3389/fspas.2023.1264226},
       adsurl = {https://ui.adsabs.harvard.edu/abs/2023FrASS..1064226H}
}

@ARTICLE{Temmer_kinem_CME_2016AN....337.1010T,
       author = {{Temmer}, M.},
        title = "{Kinematical properties of coronal mass ejections}",
      journal = {Astronomische Nachrichten},
         year = 2016,
        month = nov,
       volume = {337},
       number = {10},
        pages = {1010},
          doi = {10.1002/asna.201612425},
archivePrefix = {arXiv},
       eprint = {1603.01398},
 primaryClass = {astro-ph.SR},
       adsurl = {https://ui.adsabs.harvard.edu/abs/2016AN....337.1010T}
}

@ARTICLE{Vourlidas_CME_flux_rope_2013SoPh..284..179V,
       author = {{Vourlidas}, A. and {Lynch}, B.~J. and {Howard}, R.~A. and {Li}, Y.},
        title = "{How Many CMEs Have Flux Ropes? Deciphering the Signatures of Shocks, Flux Ropes, and Prominences in Coronagraph Observations of CMEs}",
      journal = {\solphys},
         year = 2013,
        month = may,
       volume = {284},
       number = {1},
        pages = {179-201},
          doi = {10.1007/s11207-012-0084-8},
archivePrefix = {arXiv},
       eprint = {1207.1599},
 primaryClass = {astro-ph.SR},
       adsurl = {https://ui.adsabs.harvard.edu/abs/2013SoPh..284..179V}
}

@ARTICLE{PSP_2016SSRv..204....7F,
       author = {{Fox}, N.~J. and {Velli}, M.~C. and {Bale}, S.~D. and {Decker}, R. and {Driesman}, A. and {Howard}, R.~A. and {Kasper}, J.~C. and {Kinnison}, J. and {Kusterer}, M. and {Lario}, D. and {Lockwood}, M.~K. and {McComas}, D.~J. and {Raouafi}, N.~E. and {Szabo}, A.},
        title = "{The Solar Probe Plus Mission: Humanity's First Visit to Our Star}",
      journal = {\ssr},
         year = 2016,
        month = dec,
       volume = {204},
       number = {1-4},
        pages = {7-48},
          doi = {10.1007/s11214-015-0211-6},
       adsurl = {https://ui.adsabs.harvard.edu/abs/2016SSRv..204....7F}
}

@ARTICLE{Forbes_2000JGR...10523153F,
       author = {{Forbes}, T.~G.},
        title = "{A review on the genesis of coronal mass ejections}",
      journal = {\jgr},
         year = 2000,
        month = oct,
       volume = {105},
       number = {A10},
        pages = {23153-23166},
          doi = {10.1029/2000JA000005},
       adsurl = {https://ui.adsabs.harvard.edu/abs/2000JGR...10523153F}
}

@ARTICLE{Andretta_waves_2025,
       author = {{Andretta}, V. and {Abbo}, L. and {Jerse}, G. and {Lionello}, R. and {Naletto}, G. and {Russano}, G. and {Spadaro}, D. and {Stangalini}, M. and {Susino}, R. and {Uslenghi}, M. and {Ventura}, R. and {Bemporad}, A. and {De Leo}, Y. and {Farina}, S. and {Nistic{\`o}}, G. and {Romoli}, M. and {Straus}, Th. and {Telloni}, D. and {Teriaca}, L. and {Burtovoi}, A. and {Da Deppo}, V. and {Fineschi}, S. and {Frassati}, F. and {Giarrusso}, M. and {Grimani}, C. and {Heinzel}, P. and {Landini}, F. and {Moses}, D. and {Nicolini}, G. and {Pancrazzi}, M. and {Sasso}, C.},
        title = "{First detection of acoustic-like flux in the middle solar corona}",
      journal = {\aap},
         year = 2025,
        month = sep,
       volume = {701},
          eid = {A199},
        pages = {A199},
          doi = {10.1051/0004-6361/202554034},
archivePrefix = {arXiv},
       eprint = {2507.13487},
 primaryClass = {astro-ph.SR},
       adsurl = {https://ui.adsabs.harvard.edu/abs/2025A&A...701A.199A}
}

@ARTICLE{Bemporad_CME_HR_2025,
       author = {{Bemporad}, Alessandro and {Abbo}, Lucia and {Albert}, Kinga and {Amato}, Emanuele and {Andretta}, Vincenzo and {Biondo}, Ruggero and {Burtovoi}, Aleksandr and {Calchetti}, Daniele and {Da Deppo}, Vania and {De Leo}, Yara and {Fineschi}, Silvano and {Frassati}, Federica and {Grimani}, Catia and {Jerse}, Giovanna and {Landini}, Federico and {Mancuso}, Salvatore and {Naletto}, Giampiero and {Nicolini}, Gianalfredo and {Pancrazzi}, Maurizio and {Blanco Rodr{\'\i}guez}, Julian and {Romoli}, Marco and {Russano}, Giuliana and {Sasso}, Clementina and {Spadaro}, Daniele and {Stangalini}, Marco and {Strecker}, Hanna and {Orozco Su{\'a}rez}, David and {Susino}, Roberto and {Teriaca}, Luca and {Uslenghi}, Michela and {Valori}, Gherardo},
        title = "{Discovery of Small-scale Flows in the Void of a Coronal Mass Ejection with High-cadence Images Acquired by the Metis Coronagraph on Board Solar Orbiter}",
      journal = {\apj},
         year = 2025,
        month = may,
       volume = {985},
       number = {1},
          eid = {145},
        pages = {145},
          doi = {10.3847/1538-4357/adc7ff},
       adsurl = {https://ui.adsabs.harvard.edu/abs/2025ApJ...985..145B}
}

@ARTICLE{Vourlidas_SBO_2018ApJ...861..103V,
       author = {{Vourlidas}, Angelos and {Webb}, David F.},
        title = "{Streamer-blowout Coronal Mass Ejections: Their Properties and Relation to the Coronal Magnetic Field Structure}",
      journal = {\apj},
         year = 2018,
        month = jul,
       volume = {861},
       number = {2},
          eid = {103},
        pages = {103},
          doi = {10.3847/1538-4357/aaca3e},
archivePrefix = {arXiv},
       eprint = {1806.00644},
 primaryClass = {astro-ph.SR},
       adsurl = {https://ui.adsabs.harvard.edu/abs/2018ApJ...861..103V}
}

@ARTICLE{Howard_PSP_FR_2022ApJ...936...43H,
       author = {{Howard}, Russell A. and {Stenborg}, Guillermo and {Vourlidas}, Angelos and {Gallagher}, Brendan M. and {Linton}, Mark G. and {Hess}, Phillip and {Rich}, Nathan B. and {Liewer}, Paulett C.},
        title = "{Overview of the Remote Sensing Observations from PSP Solar Encounter 10 with Perihelion at 13.3 R $_{{\ensuremath{\odot}}}$}",
      journal = {\apj},
         year = 2022,
        month = sep,
       volume = {936},
       number = {1},
          eid = {43},
        pages = {43},
          doi = {10.3847/1538-4357/ac7ff5},
archivePrefix = {arXiv},
       eprint = {2207.12175},
 primaryClass = {astro-ph.SR},
       adsurl = {https://ui.adsabs.harvard.edu/abs/2022ApJ...936...43H}
}

@ARTICLE{SECCHI_2008SSRv..136...67H,
       author = {{Howard}, R.~A. and {Moses}, J.~D. and {Vourlidas}, A. and {Newmark}, J.~S. and {Socker}, D.~G. and {Plunkett}, S.~P. and {Korendyke}, C.~M. and {Cook}, J.~W. and {Hurley}, A. and {Davila}, J.~M. and {Thompson}, W.~T. and {St Cyr}, O.~C. and {Mentzell}, E. and {Mehalick}, K. and {Lemen}, J.~R. and {Wuelser}, J.~P. and {Duncan}, D.~W. and {Tarbell}, T.~D. and {Wolfson}, C.~J. and {Moore}, A. and {Harrison}, R.~A. and {Waltham}, N.~R. and {Lang}, J. and {Davis}, C.~J. and {Eyles}, C.~J. and {Mapson-Menard}, H. and {Simnett}, G.~M. and {Halain}, J.~P. and {Defise}, J.~M. and {Mazy}, E. and {Rochus}, P. and {Mercier}, R. and {Ravet}, M.~F. and {Delmotte}, F. and {Auchere}, F. and {Delaboudiniere}, J.~P. and {Bothmer}, V. and {Deutsch}, W. and {Wang}, D. and {Rich}, N. and {Cooper}, S. and {Stephens}, V. and {Maahs}, G. and {Baugh}, R. and {McMullin}, D. and {Carter}, T.},
        title = "{Sun Earth Connection Coronal and Heliospheric Investigation (SECCHI)}",
      journal = {\ssr},
         year = 2008,
        month = apr,
       volume = {136},
       number = {1-4},
        pages = {67-115},
          doi = {10.1007/s11214-008-9341-4},
       adsurl = {https://ui.adsabs.harvard.edu/abs/2008SSRv..136...67H}
}

@ARTICLE{Thomson_surface_2012ApJ...752..130H,
       author = {{Howard}, T.~A. and {DeForest}, C.~E.},
        title = "{The Thomson Surface. I. Reality and Myth}",
      journal = {\apj},
         year = 2012,
        month = jun,
       volume = {752},
       number = {2},
          eid = {130},
        pages = {130},
          doi = {10.1088/0004-637X/752/2/130},
       adsurl = {https://ui.adsabs.harvard.edu/abs/2012ApJ...752..130H}
}

@ARTICLE{Vourlidas_CME_bright_2006ApJ...642.1216V,
       author = {{Vourlidas}, Angelos and {Howard}, Russell A.},
        title = "{The Proper Treatment of Coronal Mass Ejection Brightness: A New Methodology and Implications for Observations}",
      journal = {\apj},
         year = 2006,
        month = may,
       volume = {642},
       number = {2},
        pages = {1216-1221},
          doi = {10.1086/501122},
       adsurl = {https://ui.adsabs.harvard.edu/abs/2006ApJ...642.1216V}
}

@ARTICLE{Jess_waves_review_2023LRSP...20....1J,
       author = {{Jess}, David B. and {Jafarzadeh}, Shahin and {Keys}, Peter H. and {Stangalini}, Marco and {Verth}, Gary and {Grant}, Samuel D.~T.},
        title = "{Waves in the lower solar atmosphere: the dawn of next-generation solar telescopes}",
      journal = {Living Reviews in Solar Physics},
         year = 2023,
        month = dec,
       volume = {20},
       number = {1},
          eid = {1},
        pages = {1},
          doi = {10.1007/s41116-022-00035-6},
archivePrefix = {arXiv},
       eprint = {2212.09788},
 primaryClass = {astro-ph.SR},
       adsurl = {https://ui.adsabs.harvard.edu/abs/2023LRSP...20....1J}
}

@ARTICLE{Warmuth_waves_2015LRSP...12....3W,
       author = {{Warmuth}, Alexander},
        title = "{Large-scale Globally Propagating Coronal Waves}",
      journal = {Living Reviews in Solar Physics},
         year = 2015,
        month = dec,
       volume = {12},
       number = {1},
          eid = {3},
        pages = {3},
          doi = {10.1007/lrsp-2015-3},
       adsurl = {https://ui.adsabs.harvard.edu/abs/2015LRSP...12....3W}
}

@ARTICLE{EIT_1995SoPh..162..291D,
       author = {{Delaboudini{\`e}re}, J. -P. and {Artzner}, G.~E. and {Brunaud}, J. and {Gabriel}, A.~H. and {Hochedez}, J.~F. and {Millier}, F. and {Song}, X.~Y. and {Au}, B. and {Dere}, K.~P. and {Howard}, R.~A. and {Kreplin}, R. and {Michels}, D.~J. and {Moses}, J.~D. and {Defise}, J.~M. and {Jamar}, C. and {Rochus}, P. and {Chauvineau}, J.~P. and {Marioge}, J.~P. and {Catura}, R.~C. and {Lemen}, J.~R. and {Shing}, L. and {Stern}, R.~A. and {Gurman}, J.~B. and {Neupert}, W.~M. and {Maucherat}, A. and {Clette}, F. and {Cugnon}, P. and {Van Dessel}, E.~L.},
        title = "{EIT: Extreme-Ultraviolet Imaging Telescope for the SOHO Mission}",
      journal = {\solphys},
         year = 1995,
        month = dec,
       volume = {162},
       number = {1-2},
        pages = {291-312},
          doi = {10.1007/BF00733432},
       adsurl = {https://ui.adsabs.harvard.edu/abs/1995SoPh..162..291D}
}

@ARTICLE{EUVI_2008SSRv..136...67H,
       author = {{Howard}, R.~A. and {Moses}, J.~D. and {Vourlidas}, A. and {Newmark}, J.~S. and {Socker}, D.~G. and {Plunkett}, S.~P. and {Korendyke}, C.~M. and {Cook}, J.~W. and {Hurley}, A. and {Davila}, J.~M. and {Thompson}, W.~T. and {St Cyr}, O.~C. and {Mentzell}, E. and {Mehalick}, K. and {Lemen}, J.~R. and {Wuelser}, J.~P. and {Duncan}, D.~W. and {Tarbell}, T.~D. and {Wolfson}, C.~J. and {Moore}, A. and {Harrison}, R.~A. and {Waltham}, N.~R. and {Lang}, J. and {Davis}, C.~J. and {Eyles}, C.~J. and {Mapson-Menard}, H. and {Simnett}, G.~M. and {Halain}, J.~P. and {Defise}, J.~M. and {Mazy}, E. and {Rochus}, P. and {Mercier}, R. and {Ravet}, M.~F. and {Delmotte}, F. and {Auchere}, F. and {Delaboudiniere}, J.~P. and {Bothmer}, V. and {Deutsch}, W. and {Wang}, D. and {Rich}, N. and {Cooper}, S. and {Stephens}, V. and {Maahs}, G. and {Baugh}, R. and {McMullin}, D. and {Carter}, T.},
        title = "{Sun Earth Connection Coronal and Heliospheric Investigation (SECCHI)}",
      journal = {\ssr},
         year = 2008,
        month = apr,
       volume = {136},
       number = {1-4},
        pages = {67-115},
          doi = {10.1007/s11214-008-9341-4},
       adsurl = {https://ui.adsabs.harvard.edu/abs/2008SSRv..136...67H}
}

@ARTICLE{SDO_2012SoPh..275....3P,
       author = {{Pesnell}, W. Dean and {Thompson}, B.~J. and {Chamberlin}, P.~C.},
        title = "{The Solar Dynamics Observatory (SDO)}",
      journal = {\solphys},
         year = 2012,
        month = jan,
       volume = {275},
       number = {1-2},
        pages = {3-15},
          doi = {10.1007/s11207-011-9841-3},
       adsurl = {https://ui.adsabs.harvard.edu/abs/2012SoPh..275....3P}
}

@ARTICLE{EUI_2020A&A...642A...8R,
       author = {{Rochus}, P. and {Auch{\`e}re}, F. and {Berghmans}, D. and {Harra}, L. and {Schmutz}, W. and {Sch{\"u}hle}, U. and {Addison}, P. and {Appourchaux}, T. and {Aznar Cuadrado}, R. and {Baker}, D. and {Barbay}, J. and {Bates}, D. and {BenMoussa}, A. and {Bergmann}, M. and {Beurthe}, C. and {Borgo}, B. and {Bonte}, K. and {Bouzit}, M. and {Bradley}, L. and {B{\"u}chel}, V. and {Buchlin}, E. and {B{\"u}chner}, J. and {Cab{\'e}}, F. and {Cadiergues}, L. and {Chaigneau}, M. and {Chares}, B. and {Choque Cortez}, C. and {Coker}, P. and {Condamin}, M. and {Coumar}, S. and {Curdt}, W. and {Cutler}, J. and {Davies}, D. and {Davison}, G. and {Defise}, J. -M. and {Del Zanna}, G. and {Delmotte}, F. and {Delouille}, V. and {Dolla}, L. and {Dumesnil}, C. and {D{\"u}rig}, F. and {Enge}, R. and {Fran{\c{c}}ois}, S. and {Fourmond}, J. -J. and {Gillis}, J. -M. and {Giordanengo}, B. and {Gissot}, S. and {Green}, L.~M. and {Guerreiro}, N. and {Guilbaud}, A. and {Gyo}, M. and {Haberreiter}, M. and {Hafiz}, A. and {Hailey}, M. and {Halain}, J. -P. and {Hansotte}, J. and {Hecquet}, C. and {Heerlein}, K. and {Hellin}, M. -L. and {Hemsley}, S. and {Hermans}, A. and {Hervier}, V. and {Hochedez}, J. -F. and {Houbrechts}, Y. and {Ihsan}, K. and {Jacques}, L. and {J{\'e}r{\^o}me}, A. and {Jones}, J. and {Kahle}, M. and {Kennedy}, T. and {Klaproth}, M. and {Kolleck}, M. and {Koller}, S. and {Kotsialos}, E. and {Kraaikamp}, E. and {Langer}, P. and {Lawrenson}, A. and {Le Clech'}, J. -C. and {Lenaerts}, C. and {Liebecq}, S. and {Linder}, D. and {Long}, D.~M. and {Mampaey}, B. and {Markiewicz-Innes}, D. and {Marquet}, B. and {Marsch}, E. and {Matthews}, S. and {Mazy}, E. and {Mazzoli}, A. and {Meining}, S. and {Meltchakov}, E. and {Mercier}, R. and {Meyer}, S. and {Monecke}, M. and {Monfort}, F. and {Morinaud}, G. and {Moron}, F. and {Mountney}, L. and {M{\"u}ller}, R. and {Nicula}, B. and {Parenti}, S. and {Peter}, H. and {Pfiffner}, D. and {Philippon}, A. and {Phillips}, I. and {Plesseria}, J. -Y. and {Pylyser}, E. and {Rabecki}, F. and {Ravet-Krill}, M. -F. and {Rebellato}, J. and {Renotte}, E. and {Rodriguez}, L. and {Roose}, S. and {Rosin}, J. and {Rossi}, L. and {Roth}, P. and {Rouesnel}, F. and {Roulliay}, M. and {Rousseau}, A. and {Ruane}, K. and {Scanlan}, J. and {Schlatter}, P. and {Seaton}, D.~B. and {Silliman}, K. and {Smit}, S. and {Smith}, P.~J. and {Solanki}, S.~K. and {Spescha}, M. and {Spencer}, A. and {Stegen}, K. and {Stockman}, Y. and {Szwec}, N. and {Tamiatto}, C. and {Tandy}, J. and {Teriaca}, L. and {Theobald}, C. and {Tychon}, I. and {van Driel-Gesztelyi}, L. and {Verbeeck}, C. and {Vial}, J. -C. and {Werner}, S. and {West}, M.~J. and {Westwood}, D. and {Wiegelmann}, T. and {Willis}, G. and {Winter}, B. and {Zerr}, A. and {Zhang}, X. and {Zhukov}, A.~N.},
        title = "{The Solar Orbiter EUI instrument: The Extreme Ultraviolet Imager}",
      journal = {\aap},
         year = 2020,
        month = oct,
       volume = {642},
          eid = {A8},
        pages = {A8},
          doi = {10.1051/0004-6361/201936663},
       adsurl = {https://ui.adsabs.harvard.edu/abs/2020A&A...642A...8R}
}

@ARTICLE{SUVI_2022SpWea..2003044D,
       author = {{Darnel}, Jonathan M. and {Seaton}, Daniel B. and {Bethge}, Christian and {Rachmeler}, Laurel and {Jarvis}, Alison and {Hill}, Steven M. and {Peck}, Courtney L. and {Hughes}, J. Marcus and {Shapiro}, Jason and {Riley}, Allyssa and {Vasudevan}, Gopal and {Shing}, Lawrence and {Koener}, George and {Edwards}, Chris and {Mathur}, Dnyanesh and {Timothy}, Shelbe},
        title = "{The GOES-R Solar UltraViolet Imager}",
      journal = {Space Weather},
         year = 2022,
        month = apr,
       volume = {20},
       number = {4},
          eid = {e2022SW003044},
        pages = {e2022SW003044},
          doi = {10.1029/2022SW00304410.1002/essoar.10510311.1},
       adsurl = {https://ui.adsabs.harvard.edu/abs/2022SpWea..2003044D}
}

@ARTICLE{GCS_implementation_2011ApJS..194...33T,
       author = {{Thernisien}, A.},
        title = "{Implementation of the Graduated Cylindrical Shell Model for the Three-dimensional Reconstruction of Coronal Mass Ejections}",
      journal = {\apjs},
         year = 2011,
        month = jun,
       volume = {194},
       number = {2},
          eid = {33},
        pages = {33},
          doi = {10.1088/0067-0049/194/2/33},
       adsurl = {https://ui.adsabs.harvard.edu/abs/2011ApJS..194...33T}
}

@ARTICLE{Thernisien_2006ApJ...652..763T,
       author = {{Thernisien}, A.~F.~R. and {Howard}, R.~A. and {Vourlidas}, A.},
        title = "{Modeling of Flux Rope Coronal Mass Ejections}",
      journal = {\apj},
         year = 2006,
        month = nov,
       volume = {652},
       number = {1},
        pages = {763-773},
          doi = {10.1086/508254},
       adsurl = {https://ui.adsabs.harvard.edu/abs/2006ApJ...652..763T}
}

@ARTICLE{SolarMach_2023FrASS...958810G,
       author = {{Gieseler}, Jan and {Dresing}, Nina and {Palmroos}, Christian and {Freiherr von Forstner}, Johan L. and {Price}, Daniel J. and {Vainio}, Rami and {Kouloumvakos}, Athanasios and {Rodr{\'\i}guez-Garc{\'\i}a}, Laura and {Trotta}, Domenico and {G{\'e}not}, Vincent and {Masson}, Arnaud and {Roth}, Markus and {Veronig}, Astrid},
        title = "{Solar-MACH: An open-source tool to analyze solar magnetic connection configurations}",
      journal = {Frontiers in Astronomy and Space Sciences},
         year = 2023,
        month = feb,
       volume = {9},
          eid = {384},
        pages = {384},
          doi = {10.3389/fspas.2022.1058810},
archivePrefix = {arXiv},
       eprint = {2210.00819},
 primaryClass = {astro-ph.SR},
       adsurl = {https://ui.adsabs.harvard.edu/abs/2023FrASS...958810G}
}

@ARTICLE{JHV,
       author = {{M{\"u}ller}, D. and {Nicula}, B. and {Felix}, S. and {Verstringe}, F. and {Bourgoignie}, B. and {Csillaghy}, A. and {Berghmans}, D. and {Jiggens}, P. and {Garc{\'\i}a-Ortiz}, J.~P. and {Ireland}, J. and {Zahniy}, S. and {Fleck}, B.},
        title = "{JHelioviewer. Time-dependent 3D visualisation of solar and heliospheric data}",
      journal = {\aap},
         year = 2017,
        month = sep,
       volume = {606},
          eid = {A10},
        pages = {A10},
          doi = {10.1051/0004-6361/201730893},
archivePrefix = {arXiv},
       eprint = {1705.07628},
 primaryClass = {astro-ph.SR},
       adsurl = {https://ui.adsabs.harvard.edu/abs/2017A&A...606A..10M}
}

@ARTICLE{Sun_magn_reconn_2015NatCo...6.7598S,
       author = {{Sun}, J.~Q. and {Cheng}, X. and {Ding}, M.~D. and {Guo}, Y. and {Priest}, E.~R. and {Parnell}, C.~E. and {Edwards}, S.~J. and {Zhang}, J. and {Chen}, P.~F. and {Fang}, C.},
        title = "{Extreme ultraviolet imaging of three-dimensional magnetic reconnection in a solar eruption}",
      journal = {Nature Communications},
         year = 2015,
        month = jun,
       volume = {6},
          eid = {7598},
        pages = {7598},
          doi = {10.1038/ncomms8598},
archivePrefix = {arXiv},
       eprint = {1506.08255},
 primaryClass = {astro-ph.SR},
       adsurl = {https://ui.adsabs.harvard.edu/abs/2015NatCo...6.7598S}
}

@ARTICLE{Yashiro_CME_SOHO_catalogue_2004JGRA..109.7105Y,
       author = {{Yashiro}, S. and {Gopalswamy}, N. and {Michalek}, G. and {St. Cyr}, O.~C. and {Plunkett}, S.~P. and {Rich}, N.~B. and {Howard}, R.~A.},
        title = "{A catalog of white light coronal mass ejections observed by the SOHO spacecraft}",
      journal = {Journal of Geophysical Research (Space Physics)},
         year = 2004,
        month = jul,
       volume = {109},
       number = {A7},
          eid = {A07105},
        pages = {A07105},
          doi = {10.1029/2003JA010282},
       adsurl = {https://ui.adsabs.harvard.edu/abs/2004JGRA..109.7105Y}
}

@ARTICLE{wavelets_1998BAMS...79...61T,
       author = {{Torrence}, Christopher and {Compo}, Gilbert P.},
        title = "{A Practical Guide to Wavelet Analysis.}",
      journal = {Bulletin of the American Meteorological Society},
         year = 1998,
        month = jan,
       volume = {79},
       number = {1},
        pages = {61-78},
          doi = {10.1175/1520-0477(1998)079<0061:APGTWA>2.0.CO;2},
       adsurl = {https://ui.adsabs.harvard.edu/abs/1998BAMS...79...61T}
}

@ARTICLE{EMD_1998RSPSA.454..903H,
       author = {{Huang}, N.~E. and {Shen}, Z. and {Long}, S.~R. and {Wu}, M.~C. and {Shih}, H.~H. and {Zheng}, Q. and {Yen}, N. -C. and {Tung}, C.~C. and {Liu}, H.~H.},
        title = "{The empirical mode decomposition and the Hilbert spectrum for nonlinear and non-stationary time series analysis}",
      journal = {Proceedings of the Royal Society of London Series A},
         year = 1998,
        month = mar,
       volume = {454},
       number = {1971},
        pages = {903-998},
          doi = {10.1098/rspa.1998.0193},
       adsurl = {https://ui.adsabs.harvard.edu/abs/1998RSPSA.454..903H}
}

@ARTICLE{Robbrecht_stealth_2009ApJ...701..283R,
       author = {{Robbrecht}, Eva and {Patsourakos}, Spiros and {Vourlidas}, Angelos},
        title = "{No Trace Left Behind: STEREO Observation of a Coronal Mass Ejection Without Low Coronal Signatures}",
      journal = {\apj},
         year = 2009,
        month = aug,
       volume = {701},
       number = {1},
        pages = {283-291},
          doi = {10.1088/0004-637X/701/1/283},
archivePrefix = {arXiv},
       eprint = {0905.2583},
 primaryClass = {astro-ph.SR},
       adsurl = {https://ui.adsabs.harvard.edu/abs/2009ApJ...701..283R}
}

@ARTICLE{Lynch_stealth_2016JGRA..12110677L,
       author = {{Lynch}, B.~J. and {Masson}, S. and {Li}, Y. and {DeVore}, C.~R. and {Luhmann}, J.~G. and {Antiochos}, S.~K. and {Fisher}, G.~H.},
        title = "{A model for stealth coronal mass ejections}",
      journal = {Journal of Geophysical Research (Space Physics)},
         year = 2016,
        month = nov,
       volume = {121},
       number = {11},
        pages = {10677-10697},
          doi = {10.1002/2016JA023432},
archivePrefix = {arXiv},
       eprint = {1612.08323},
 primaryClass = {astro-ph.SR},
       adsurl = {https://ui.adsabs.harvard.edu/abs/2016JGRA..12110677L}
}

@ARTICLE{Howard_stealth_2013SoPh..285..269H,
       author = {{Howard}, Timothy A. and {Harrison}, Richard A.},
        title = "{Stealth Coronal Mass Ejections: A Perspective}",
      journal = {\solphys},
         year = 2013,
        month = jul,
       volume = {285},
       number = {1-2},
        pages = {269-280},
          doi = {10.1007/s11207-012-0217-0},
       adsurl = {https://ui.adsabs.harvard.edu/abs/2013SoPh..285..269H}
}

@ARTICLE{Foullon_KH_2011ApJ...729L...8F,
       author = {{Foullon}, Claire and {Verwichte}, Erwin and {Nakariakov}, Valery M. and {Nykyri}, Katariina and {Farrugia}, Charles J.},
        title = "{Magnetic Kelvin-Helmholtz Instability at the Sun}",
      journal = {\apjl},
         year = 2011,
        month = mar,
       volume = {729},
       number = {1},
          eid = {L8},
        pages = {L8},
          doi = {10.1088/2041-8205/729/1/L8},
       adsurl = {https://ui.adsabs.harvard.edu/abs/2011ApJ...729L...8F}
}

@ARTICLE{Paouris_KH_PSP_2024ApJ...964..139P,
       author = {{Paouris}, Evangelos and {Stenborg}, Guillermo and {Linton}, Mark G. and {Vourlidas}, Angelos and {Howard}, Russell A. and {Raouafi}, Nour E.},
        title = "{First Direct Imaging of a Kelvin{\textendash}Helmholtz Instability by PSP/WISPR}",
      journal = {\apj},
         year = 2024,
        month = apr,
       volume = {964},
       number = {2},
          eid = {139},
        pages = {139},
          doi = {10.3847/1538-4357/ad2208},
       adsurl = {https://ui.adsabs.harvard.edu/abs/2024ApJ...964..139P}
}

@BOOK{Chandrasekhar,
       author = {{Chandrasekhar}, S.},
        title = "{Hydrodynamic and Hydromagnetic Stability (Oxford: Clarendon)}",
         year = 1961
}

@ARTICLE{Syntelis_KH_sim_2019ApJ...884L...4S,
       author = {{Syntelis}, P. and {Antolin}, P.},
        title = "{Kelvin-Helmholtz Instability and Alfv{\'e}nic Vortex Shedding in Solar Eruptions}",
      journal = {\apjl},
         year = 2019,
        month = oct,
       volume = {884},
       number = {1},
          eid = {L4},
        pages = {L4},
          doi = {10.3847/2041-8213/ab44ab},
archivePrefix = {arXiv},
       eprint = {1909.05716},
 primaryClass = {astro-ph.SR},
       adsurl = {https://ui.adsabs.harvard.edu/abs/2019ApJ...884L...4S}
}

@ARTICLE{Liu_QPF_2012ApJ...753...52L,
       author = {{Liu}, Wei and {Ofman}, Leon and {Nitta}, Nariaki V. and {Aschwanden}, Markus J. and {Schrijver}, Carolus J. and {Title}, Alan M. and {Tarbell}, Theodore D.},
        title = "{Quasi-periodic Fast-mode Wave Trains within a Global EUV Wave and Sequential Transverse Oscillations Detected by SDO/AIA}",
      journal = {\apj},
         year = 2012,
        month = jul,
       volume = {753},
       number = {1},
          eid = {52},
        pages = {52},
          doi = {10.1088/0004-637X/753/1/52},
archivePrefix = {arXiv},
       eprint = {1204.5470},
 primaryClass = {astro-ph.SR},
       adsurl = {https://ui.adsabs.harvard.edu/abs/2012ApJ...753...52L}
}

@ARTICLE{Shen_QPF_2022SoPh..297...20S,
       author = {{Shen}, Yuandeng and {Zhou}, Xinping and {Duan}, Yadan and {Tang}, Zehao and {Zhou}, Chengrui and {Tan}, Song},
        title = "{Coronal Quasi-periodic Fast-mode Propagating Wave Trains}",
      journal = {\solphys},
         year = 2022,
        month = feb,
       volume = {297},
       number = {2},
          eid = {20},
        pages = {20},
          doi = {10.1007/s11207-022-01953-2},
archivePrefix = {arXiv},
       eprint = {2112.14959},
 primaryClass = {astro-ph.SR},
       adsurl = {https://ui.adsabs.harvard.edu/abs/2022SoPh..297...20S}
}

@ARTICLE{Shen_QFP_ex_2022A&A...665A..51S,
       author = {{Shen}, Yuandeng and {Yao}, Surui and {Tang}, Zehao and {Zhou}, Xinping and {Qu}, Zhining and {Duan}, Yadan and {Zhou}, Chengrui and {Tan}, Song},
        title = "{White-light QFP wave train and the associated failed breakout eruption}",
      journal = {\aap},
         year = 2022,
        month = sep,
       volume = {665},
          eid = {A51},
        pages = {A51},
          doi = {10.1051/0004-6361/202243924},
archivePrefix = {arXiv},
       eprint = {2207.08110},
 primaryClass = {astro-ph.SR},
       adsurl = {https://ui.adsabs.harvard.edu/abs/2022A&A...665A..51S}
}

@ARTICLE{Patsourakos_QFP_Ex_2010A&A...522A.100P,
       author = {{Patsourakos}, S. and {Vourlidas}, A. and {Kliem}, B.},
        title = "{Toward understanding the early stages of an impulsively accelerated coronal mass ejection. SECCHI observations}",
      journal = {\aap},
         year = 2010,
        month = nov,
       volume = {522},
          eid = {A100},
        pages = {A100},
          doi = {10.1051/0004-6361/200913599},
archivePrefix = {arXiv},
       eprint = {1008.1171},
 primaryClass = {astro-ph.SR},
       adsurl = {https://ui.adsabs.harvard.edu/abs/2010A&A...522A.100P}
}

@ARTICLE{Jess_wave_review_2023LRSP...20....1J,
       author = {{Jess}, David B. and {Jafarzadeh}, Shahin and {Keys}, Peter H. and {Stangalini}, Marco and {Verth}, Gary and {Grant}, Samuel D.~T.},
        title = "{Waves in the lower solar atmosphere: the dawn of next-generation solar telescopes}",
      journal = {Living Reviews in Solar Physics},
         year = 2023,
        month = dec,
       volume = {20},
       number = {1},
          eid = {1},
        pages = {1},
          doi = {10.1007/s41116-022-00035-6},
archivePrefix = {arXiv},
       eprint = {2212.09788},
 primaryClass = {astro-ph.SR},
       adsurl = {https://ui.adsabs.harvard.edu/abs/2023LRSP...20....1J}
}

@misc{euidatarelease6,
  author       = {{Kraaikamp}, E. and {Gissot}, S. and  {Stegen}, K.  and {Mampaey}, B. and {Verbeeck}, F.  and  {Auch{\`e}re}, F. and {Berghmans}, D. },
  title        = {SolO/EUI Data Release 6.0 2023-01},
  howpublished = {https://doi.org/10.24414/z818-4163},
  month        = {January},
  year         = {2023},
  note         = {Published by Royal Observatory of Belgium (ROB)}
}

@ARTICLE{WOW_filter_2023A&A...670A..66A,
       author = {{Auch{\`e}re}, F. and {Soubri{\'e}}, E. and {Pelouze}, G. and {Buchlin}, {\'E}.},
        title = "{Image enhancement with wavelet-optimized whitening}",
      journal = {\aap},
         year = 2023,
        month = feb,
       volume = {670},
          eid = {A66},
        pages = {A66},
          doi = {10.1051/0004-6361/202245345},
archivePrefix = {arXiv},
       eprint = {2212.10134},
 primaryClass = {astro-ph.IM},
       adsurl = {https://ui.adsabs.harvard.edu/abs/2023A&A...670A..66A}
}

@ARTICLE{Mostl_KHi_2013ApJ...766L..12M,
       author = {{M{\"o}stl}, U.~V. and {Temmer}, M. and {Veronig}, A.~M.},
        title = "{The Kelvin-Helmholtz Instability at Coronal Mass Ejection Boundaries in the Solar Corona: Observations and 2.5D MHD Simulations}",
      journal = {\apjl},
         year = 2013,
        month = mar,
       volume = {766},
       number = {1},
          eid = {L12},
        pages = {L12},
          doi = {10.1088/2041-8205/766/1/L12},
archivePrefix = {arXiv},
       eprint = {1304.5884},
 primaryClass = {astro-ph.SR},
       adsurl = {https://ui.adsabs.harvard.edu/abs/2013ApJ...766L..12M}
}

@ARTICLE{SDO_HMI_2012SoPh..275..207S,
       author = {{Scherrer}, P.~H. and {Schou}, J. and {Bush}, R.~I. and {Kosovichev}, A.~G. and {Bogart}, R.~S. and {Hoeksema}, J.~T. and {Liu}, Y. and {Duvall}, T.~L. and {Zhao}, J. and {Title}, A.~M. and {Schrijver}, C.~J. and {Tarbell}, T.~D. and {Tomczyk}, S.},
        title = "{The Helioseismic and Magnetic Imager (HMI) Investigation for the Solar Dynamics Observatory (SDO)}",
      journal = {\solphys},
         year = 2012,
        month = jan,
       volume = {275},
       number = {1-2},
        pages = {207-227},
          doi = {10.1007/s11207-011-9834-2},
       adsurl = {https://ui.adsabs.harvard.edu/abs/2012SoPh..275..207S}
}

@ARTICLE{Palmerio_2021FrASS...8..109P,
       author = {{Palmerio}, Erika and {Nitta}, Nariaki V. and {Mulligan}, Tamitha and {Mierla}, Marilena and {O'Kane}, Jennifer and {Richardson}, Ian G. and {Sinha}, Suvadip and {Srivastava}, Nandita and {Yardley}, Stephanie L. and {Zhukov}, Andrei N.},
        title = "{Investigating Remote-sensing Techniques to Reveal Stealth Coronal Mass Ejections}",
      journal = {Frontiers in Astronomy and Space Sciences},
         year = 2021,
        month = jul,
       volume = {8},
          eid = {695966},
        pages = {695966},
          doi = {10.3389/fspas.2021.695966},
archivePrefix = {arXiv},
       eprint = {2106.07571},
 primaryClass = {astro-ph.SR},
       adsurl = {https://ui.adsabs.harvard.edu/abs/2021FrASS...8..109P}
}

@ARTICLE{Ofman_KHi_CME_2026ApJ...997L..28O,
       author = {{Ofman}, Leon and {Khabarova}, Olga and {Kwon}, Ryun-Yong and {Yogesh} and {Heifetz}, Eyal and {Nykyri}, Katariina},
        title = "{Observation of Large-scale Kelvin─Helmholtz Instability Wave Driven by a Coronal Mass Ejection}",
      journal = {\apjl},
         year = 2026,
        month = jan,
       volume = {997},
       number = {1},
          eid = {L28},
        pages = {L28},
          doi = {10.3847/2041-8213/ae3136},
archivePrefix = {arXiv},
       eprint = {2512.19942},
 primaryClass = {astro-ph.SR},
       adsurl = {https://ui.adsabs.harvard.edu/abs/2026ApJ...997L..28O}
}

@ARTICLE{Sheeley_Wang_inflows_2002ApJ...579..874S,
       author = {{Sheeley}, Jr., N.~R. and {Wang}, Y.-M.},
        title = "{Characteristics of Coronal Inflows}",
      journal = {\apj},
         year = 2002,
        month = nov,
       volume = {579},
       number = {2},
        pages = {874-887},
          doi = {10.1086/342923},
       adsurl = {https://ui.adsabs.harvard.edu/abs/2002ApJ...579..874S}
}

@ARTICLE{Tripathi_magn_reconn_2006A&A...453.1111T,
       author = {{Tripathi}, D. and {Isobe}, H. and {Mason}, H.~E.},
        title = "{On the propagation of brightening after filament/prominence eruptions, as seen by SoHO-EIT}",
      journal = {\aap},
         year = 2006,
        month = jul,
       volume = {453},
       number = {3},
        pages = {1111-1116},
          doi = {10.1051/0004-6361:20064993},
       adsurl = {https://ui.adsabs.harvard.edu/abs/2006A&A...453.1111T}
}

@ARTICLE{Tripathi_downflow_2007A&A...472..633T,
       author = {{Tripathi}, D. and {Solanki}, S.~K. and {Mason}, H.~E. and {Webb}, D.~F.},
        title = "{A bright coronal downflow seen in multi-wavelength observations: evidence of a bifurcating flux-rope?}",
      journal = {\aap},
         year = 2007,
        month = sep,
       volume = {472},
       number = {2},
        pages = {633-642},
          doi = {10.1051/0004-6361:20077707},
archivePrefix = {arXiv},
       eprint = {0802.3616},
 primaryClass = {astro-ph},
       adsurl = {https://ui.adsabs.harvard.edu/abs/2007A&A...472..633T}
}

@ARTICLE{Hu_sim_QFP_2024ApJ...962...42H,
       author = {{Hu}, Jialiang and {Ye}, Jing and {Chen}, Yuhao and {Mei}, Zhixing and {Tang}, Zehao and {Lin}, Jun},
        title = "{Excitation of Quasiperiodic Fast-propagating Waves in the Early Stage of the Solar Eruption}",
      journal = {\apj},
         year = 2024,
        month = feb,
       volume = {962},
       number = {1},
          eid = {42},
        pages = {42},
          doi = {10.3847/1538-4357/ad1993},
archivePrefix = {arXiv},
       eprint = {2312.17048},
 primaryClass = {astro-ph.SR},
       adsurl = {https://ui.adsabs.harvard.edu/abs/2024ApJ...962...42H}
}

@ARTICLE{Cappello_2024A&A...688A.162C,
       author = {{Cappello}, G.~M. and {Temmer}, M. and {Vourlidas}, A. and {Braga}, C. and {Liewer}, P.~C. and {Qiu}, J. and {Stenborg}, G. and {Kouloumvakos}, A. and {Veronig}, A.~M. and {Bothmer}, V.},
        title = "{Internal magnetic field structures observed by PSP/WISPR in a filament-related coronal mass ejection}",
      journal = {\aap},
         year = 2024,
        month = aug,
       volume = {688},
          eid = {A162},
        pages = {A162},
          doi = {10.1051/0004-6361/202449613},
archivePrefix = {arXiv},
       eprint = {2402.14682},
 primaryClass = {astro-ph.SR},
       adsurl = {https://ui.adsabs.harvard.edu/abs/2024A&A...688A.162C}
}

\begin{appendix} 

\section{3D geometric reconstruction of the CME}
\label{appendix:GCS}

To recover the 3D geometry of the CME, we used the Graduated Cylindrical Shell (GCS) model \citep{Thernisien_2006ApJ...652..763T,GCS_implementation_2011ApJS..194...33T}. It is an empirical 3D reconstruction technique, widely used to study the morphology, position and kinematics of CMEs when they are approximately less than ten solar radii from the Sun, by representing their magnetic flux rope structure.  
This reconstruction process combines simultaneous observations from multiple coronagraphs in addition with a disk imager (possible in the SolarSoftWare environment available for the NV5/IDL software), each observing the same event from a different viewpoint. 
The plane of the sky of SolO is approximately 73 degrees relative to that of STEREO-A and 45 degrees relative to SOHO and Earth (see Fig.\,\ref{fig:orbit_plot}). 

\begin{figure}
   \centering
   \includegraphics[clip, trim=4.7cm 13cm 4.5cm 3cm,width=5.6cm]{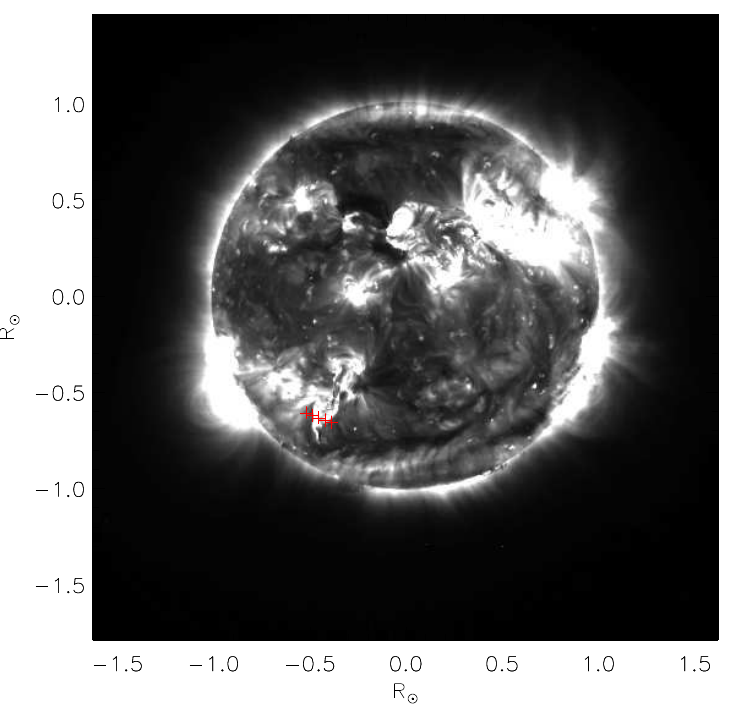}
   \includegraphics[clip, trim=5cm 13cm 4.5cm 3cm,width=5.4cm]{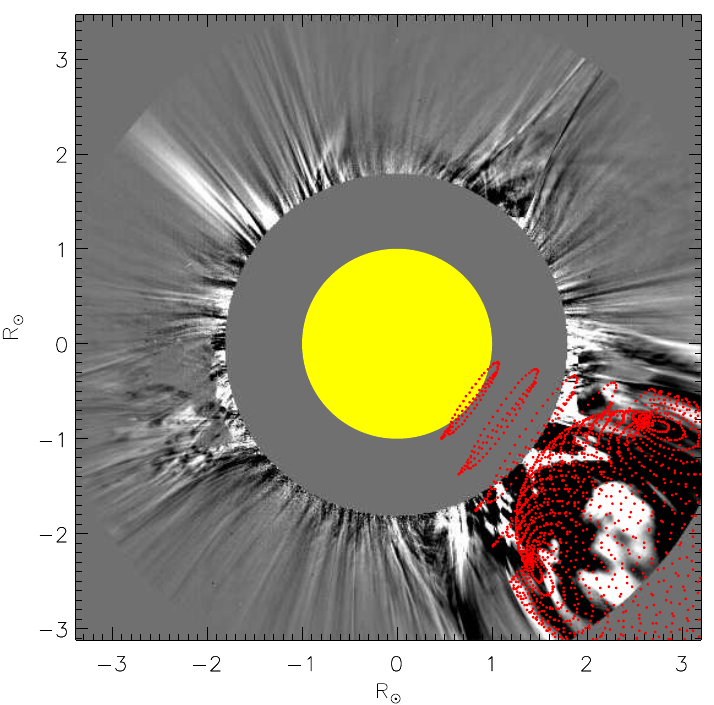}
   \includegraphics[clip, trim=5cm 13cm 4.5cm 3cm,width=5.4cm]{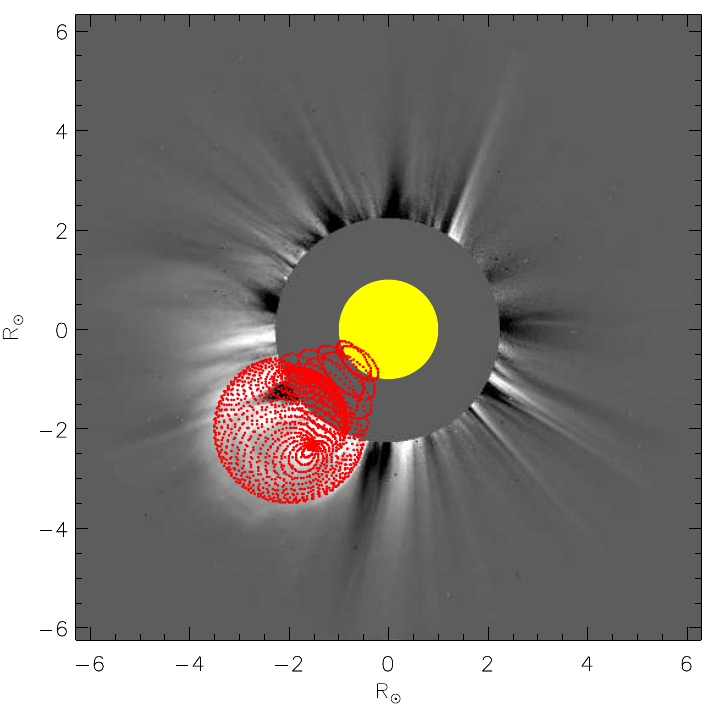}
   \includegraphics[clip, trim=4.8cm 13cm 4.5cm 3cm,width=5.5cm]{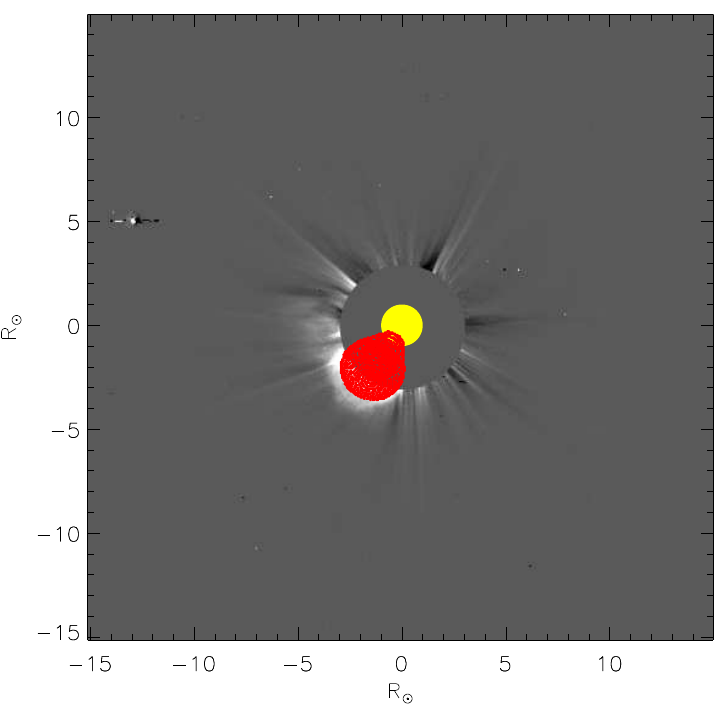}
   \caption{Graduated cylindrical shell 3D reconstruction of the CME. Near-simultaneous images from the four instruments used are shown. From left to right: STEREO/SECCHI EUVI-A 195\,$\r{A}$ (04:45:00), Metis running difference (04:36:01), SOHO/LASCO-C2 (04:40:08) and STEREO/SECCHI COR2-A (04:38:30).}
              \label{fig:GCS}%
    \end{figure}

As illustrated in Fig.\,\ref{fig:GCS}, a cylindrical shell configuration (often referred to as 'croissant' model) is employed to represent the magnetic flux rope. 
The model parameters were adjusted to ensure that the red mesh aligns with the observed CME structure in all coronagraph images, including the footpoints' location with respect to the source region on the disk. The two outer red crosses in the first panel of Fig.\,\ref{fig:GCS} mark the intersection of the two GCS conic legs axes with the solar surface, while the three inner crosses trace the connecting arc as a visual guide.

The geometrical parameters derived from the red-mesh fitting are: Carrington longitude 339$^\circ$ and latitude -32$^\circ$; height 4.6\,R$_\odot$; tilt angle -32$^\circ$; aspect ratio 0.4 and half angle 4.2$^\circ$.

\end{appendix}

\end{document}